%% file: main.tex
\documentclass[a4paper, 12pt]{article} 
\usepackage[utf8]{inputenc}
\usepackage[a4paper, width=170mm,top=25mm,bottom=15mm]{geometry}

\usepackage{titlesec}
\usepackage{placeins}

\titleformat{\section}
  {\normalfont\Large\bfseries}{\thesection}{1em}{}
  [\clearpage]

\usepackage{graphicx}    
\graphicspath{{figs/}} 

\usepackage{float}
\usepackage{caption}
\usepackage{subcaption}  
\usepackage{amsmath}  
\usepackage{amsfonts}
\usepackage{amssymb}   
\usepackage{amsthm}
\usepackage{thmtools}
\usepackage{indentfirst} 
\usepackage{url}
\usepackage{hyperref}
\usepackage{wrapfig}
\usepackage[dvipsnames]{xcolor}
\usepackage{titlesec, color}
\usepackage{ifthen}
\usepackage{etoolbox}

\input{config/settings.tex}

\title{Predicting User Grasp Intentions in Virtual Reality}
\subtitle{}
\author{
Linghao Zeng \\
\vspace{0.3cm}
\small M2 Intelligence Artificielle, Systèmes, Données (IASD)
}
\date{11/09/2024}


\logo{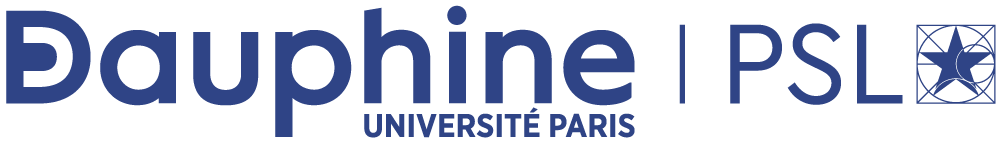}
\secondlogo{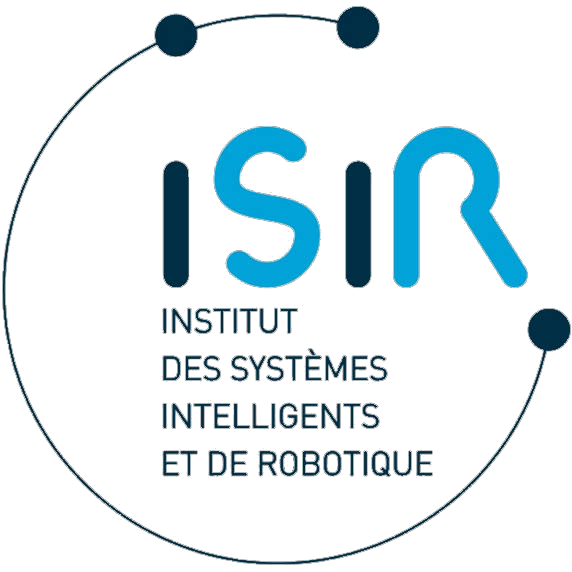}  

\firstsupervisor{Gilles Bailly}
\firstsupervisorrole{Sorbonne Université, Paris, France}

\secondsupervisor{Vincent Guigue}
\secondsupervisorrole{AgroParisTech, Paris, France}

\begin{document}

\input{config/titlepage.tex}

\newgeometry{width=140mm,top=40mm,bottom=40mm}

\section*{Abstract}
\noindent Predicting user intentions in virtual reality (VR) is crucial for creating immersive experiences, particularly in tasks involving complex grasping motions where accurate haptic feedback is essential. In this work, we leverage time-series data from hand movements to evaluate both classification and regression approaches across 810 trials with varied object types, sizes, and manipulations. Our findings reveal that classification models struggle to generalize across users, leading to inconsistent performance. In contrast, regression-based approaches, particularly those using Long Short Term Memory (LSTM) networks, demonstrate more robust performance, with timing errors within 0.25 seconds and distance errors around 5-20 cm in the critical two-second window before a grasp. Despite these improvements, predicting precise hand postures remains challenging. Through a comprehensive analysis of user variability and model interpretability, we explore why certain models fail and how regression models better accommodate the dynamic and complex nature of user behavior in VR. Our results underscore the potential of machine learning models to enhance VR interactions, particularly through adaptive haptic feedback, and lay the groundwork for future advancements in real-time prediction of user actions in VR.

\vspace{1em}
\noindent \textbf{Keywords:} Virtual Reality, Grasp Prediction, Machine Learning
\newgeometry{width=140mm,top=25mm,bottom=25mm}
\tableofcontents

\newgeometry{width=170mm,top=35mm,bottom=35mm}
\pagestyle{plain}

\section{Introduction}\label{chp:intro}

\subsection{Context: Bare-Hand Interactions in VR}
In virtual reality (VR), the future of haptic interaction is fundamentally linked to the freedom of users to interact with virtual environments using their bare hands \cite{10.1145/3379337.3415891,  10.1145/3450522.3451323}. Unlike traditional controller-based systems, where interaction is limited by buttons and predefined actions, bare-hand interaction offers a richer, more intuitive experience. Users can perform various manipulations, engaging with virtual objects as they would in the real world, experiencing tactile and kinesthetic feedback.

However, despite the advancements in VR technology, most systems today still rely heavily on controllers, which limit the scope and naturalness of interaction. This reliance contrasts with the ideal of "natural" VR interactions, which should allow users to manipulate objects freely without the constraints imposed by wearables, controllers, or exoskeletons. Recent developments, such as redirection techniques like Haptic Retargeting \cite{10.1145/2858036.2858226} and Robotic Graphics \cite{380761}, represent significant steps forward in overcoming these limitations. Haptic Retargeting modifies the user's virtual hand trajectory or grasp size to ensure alignment with physical props during interaction with virtual objects. Alternatively, Robotic Graphics uses robots to dynamically move or reshape physical props, ensuring that they accurately correspond to the virtual environment. These approaches are crucial for enabling seamless and immersive interactions with objects of varying sizes and shapes in VR.

\begin{figure}[h] \centering \includegraphics[width=\linewidth]{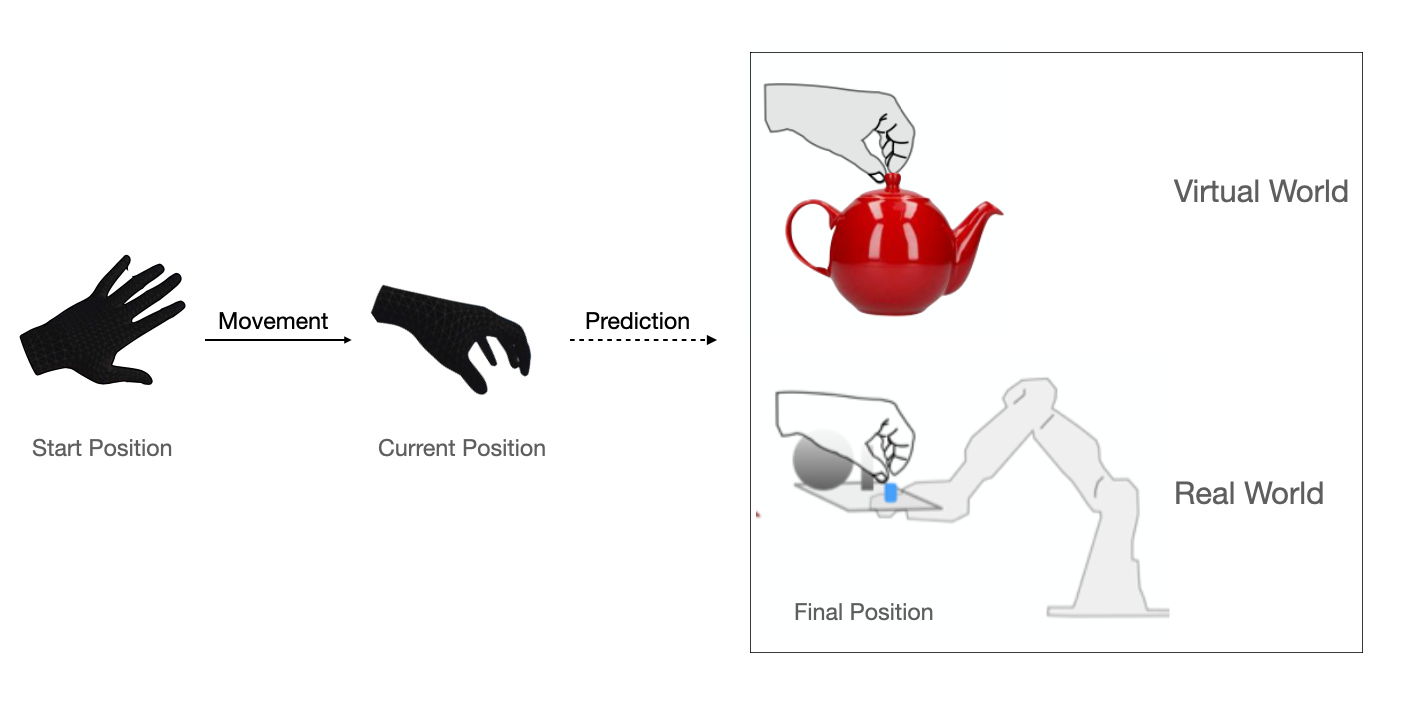} \caption{Predicting the final grasp position: As a user moves their hand towards an object, the system needs to analyze the past movement and predict the final grasp configuration. Accurate prediction allows a robotic arm to adjust the physical prop in the real world to correspond with the virtual object in the VR environment, ensuring a seamless interaction experience.} \label{fig:problem} \end{figure}

\subsection{Problem: Challenges of Real-Time User Intention Prediction}
The ability to accurately predict user intentions in real-time is critical for creating intuitive and immersive VR experiences. Specifically, predicting \textit{when}, \textit{where}, and \textit{how} users intend to interact with objects is essential for delivering timely and appropriate haptic feedback, which significantly enhances immersion \cite{9535175, 10.1145/3379337.3415870}. By anticipating user actions, VR systems can preload haptic responses, synchronize virtual objects with physical props, and dynamically adjust the environment. This reduces latency, increases system responsiveness, and ultimately creates a more seamless and intuitive VR experience \cite{10.1145/3379337.3415870, DILUCA2011245, 7247759}.

As illustrated in Figure \ref{fig:problem}, when a user begins to move their hand towards an object, accurately predicting the final grasp configuration is crucial. The system must analyze past movement data and predict the specific part of the object that the user will interact with. This prediction allows the corresponding physical prop in the real world to be adjusted in advance, ensuring that the user's interaction is mirrored accurately in both the virtual and physical environments. However, this process is complex due to the potential for multiple grasp configurations, which depend on the user’s intent and the specific manipulation of the object.

\begin{figure}[h] \centering \includegraphics[width=0.8\linewidth]{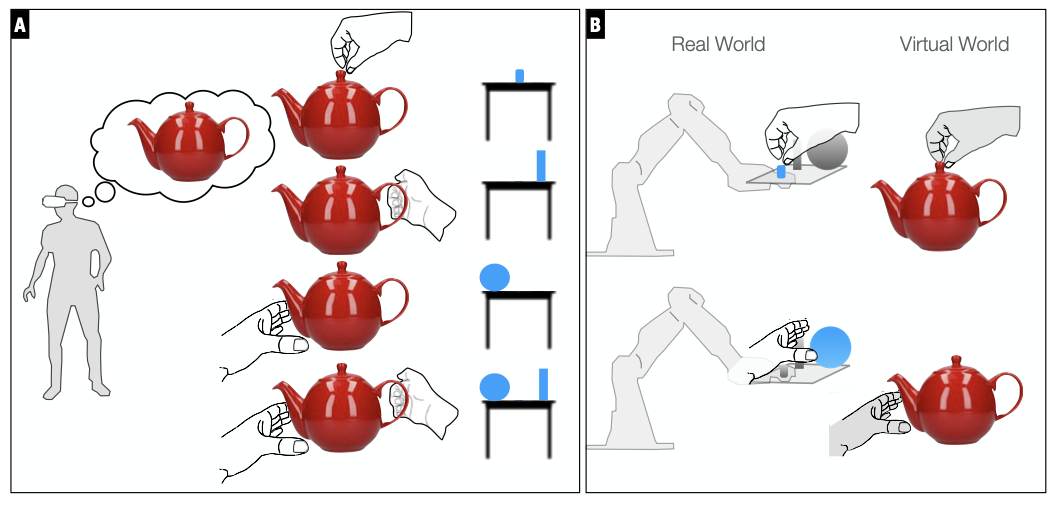} \caption{(A) A user wants to interact with a teapot—a complex-shaped object. Depending on their manipulation, the user may interact with different parts of the object (e.g., single-handed from the top or two-handed from the sides). (B) A robotic arm in the real world adapts and aligns the physical prop (e.g., a small cylinder mimicking the top of the teapot) to correspond with the virtual object, ensuring that the user’s interaction with the virtual teapot is mirrored accurately by the physical prop.} \label{fig:tea} \end{figure}

For example, in the scenario depicted in Figure \ref{fig:tea}(A), a user might grasp a teapot in various ways—by its handle, its lid, or from the side. The VR system's ability to predict the specific grasp configuration in advance allows it to synchronize the physical props with the virtual environment effectively. As shown in Figure \ref{fig:tea}(B), once the system predicts the intended grasp, a robotic arm can preemptively position the physical prop to match the virtual teapot's interaction point. This alignment is crucial for delivering a synchronized and realistic interaction experience.

\subsection{Objectives: A User-Centered Approach to Interaction} User interactions in VR are inherently complex and varied, influenced by individual preferences, task demands, and environmental cues \cite{10.3389/fnhum.2012.00117, 10.1145/3544549.3585773}. To address these challenges, our objective is to develop dynamic, real-time predictions of user intentions, facilitating a more flexible, user-centered approach to VR interaction. This approach aims to adapt to the unique ways in which each user engages with virtual environments, ensuring that the system can accommodate diverse behaviors and preferences.

By focusing on a user-centered approach, we aim to enhance the accuracy of predictions, improve the responsiveness of haptic feedback systems, and ultimately create a more immersive and intuitive VR experience. This approach not only addresses the variability in user behavior but also ensures that interactions remain seamless and natural, regardless of the user’s individual style or the specific task at hand.

\subsection{Approach: Predicting User Intentions with Machine Learning on Time-series Data} 
Our approach utilizes machine learning techniques to analyze time-series data from hand tracking systems, enabling the prediction of user actions before they occur. Hand movements during object interactions are inherently sequential, with each movement building upon the previous one. This temporal structure is crucial for accurately predicting user intentions, as the configuration of fingers and the dynamics of hand movements during reach-to-grasp actions offer valuable insights into the user’s intended interaction well before contact is made. Ansuini et al. \cite{ansuini2006effects, ansuini2015predicting} provided foundational insights into the temporal dynamics of joint movements, illustrating how wrist velocity, grasp aperture, and task end-goals influence hand shaping during reach-to-grasp motions. Cavallo et al. \cite{Cavallo2016} extended this understanding by demonstrating that subtle variations in movement kinematics can reveal underlying user intentions. Building on these analyses, Furmanek et al. \cite{Furmanek2019} examined reach-to-grasp coordination in both physical and haptic-free virtual environments, and Valkov et al. \cite{10.1145/3544549.3585773} applied these principles by leveraging phalange vectors and hand velocity with LSTM networks to predict which object a user would grasp. 

We conducted experiments across 810 trials involving 6 users, with variations in object types, sizes, and manipulations, to evaluate our approach. Our analysis addressed the challenge from both classification and regression perspectives, with a particular emphasis on how regression models could better accommodate the inherent variability in user behavior. The study revealed that classification models often struggle with the complexity and unpredictability of user interactions in virtual environments, whereas regression models demonstrated a more robust ability to adapt to these variations. Unlike previous work that primarily focused on predicting which object a user would interact with \cite{10.1145/3379337.3415891, 10.1145/3544549.3585773, 9417717}, our approach using regression models explores within-object predictions. This allows for a more nuanced understanding of user intentions by capturing subtle variations in hand positioning and motion as the action unfolds over time. By leveraging time-series data from hand tracking, regression models provided more precise and adaptable predictions of continuous outcomes, such as the exact position and timing of grasps. These findings underscore the potential of regression-based models to better accommodate the diverse and dynamic nature of user behavior in VR, highlighting their significance in developing more accurate and responsive virtual interaction systems.

The main contributions of this work are: 
\begin{enumerate} 
\item \textbf{Regression-Based Framework for User Intention Prediction:} We developed a regression approach that more effectively predicts continuous grasp posture, position, and timing in VR environments. 
\item \textbf{Analysis and Interpretability of User Variability:} Through in-depth analysis, we examined the impact of user variability on prediction model performance, offering insights into why certain models struggle to generalize across different users. This analysis highlights the limitations and potential failure points of both classification and regression models in handling the complexity of user behavior in VR. 
\item \textbf{Visualization for Model Understanding:} We utilized visualizations to enhance the understanding of model behavior, providing a detailed view of how different models predict grasp postures, positions and timings, and identifying the causes of discrepancies. \end{enumerate}

\section{Related Work}
\subsection{Bare-Hand Interactions in VR}
Bare-hand interactions in VR have been explored as a way to create more immersive and natural user experiences. However, most existing systems focus on simple tasks like target acquisition and object selection \cite{7781770, 10.1145/3411764.3445193, 9417717}, without addressing the complexities of human grasping. This limits direct manipulation, where users should be able to interact with virtual objects as naturally as they would in the real world.

A critical challenge in achieving natural bare-hand interactions is the absence of adequate haptic feedback. Haptic cues are essential for grasping as they provide crucial information about an object's properties, such as texture, weight, and resistance. Without these cues, users often struggle with precise tasks, particularly in VR environments where depth perception can be distorted \cite{bozzacchi2015lack, 10.1145/3366550.3372248}. Although systems like Oculus attempt to replicate real-hand movements, they fall short in supporting multi-finger grasping and delivering realistic haptic feedback, thus limiting the quality of interaction.

Various solutions have been proposed to bridge this gap. Modular controllers and wearables, such as pop-up props or pin-based arrays, attempt to simulate basic object shapes \cite{10.1145/3242587.3242628, 10.1145/3313831.3376358}. Exoskeletons and other wearable devices provide haptic feedback by constraining the user's fingers according to the virtual object boundaries \cite{7223325, 10.1145/3313831.3376470}. However, these approaches often have limitations because discrepancies between expected and actual feedback can disrupt immersion, and the hardware can be cumbersome for users \cite{wexelblat1993virtual}.

More promising are approaches that prioritize direct manipulation through passive props, which can offer more natural haptic experiences \cite{10.5555/933178}. These methods map physical objects to the virtual environment, allowing users to perform a variety of grasp types as they would in real life. However, these solutions rely heavily on the ability to predict user intentions in real-time, enabling the system to provide appropriate physical feedback through technologies such as robotic interfaces or redirection techniques \cite{10.1145/3379337.3415891, 10.1145/3379337.3415870}.

Our work seeks to address these challenges by predicting user grasp intentions in VR without relying on cumbersome wearables or controllers. By anticipating user actions, we aim to deliver timely and contextually appropriate haptic feedback, enhancing the realism and immersion of bare-hand interactions in VR environments. This approach contributes to more intuitive and engaging VR experiences, pushing the boundaries of what is possible with bare-hand interaction in virtual spaces.

\subsection{The Dynamics of Human Motion: From Reach to Grasp}
\begin{figure}
    \centering
    \includegraphics[width=0.8\linewidth]{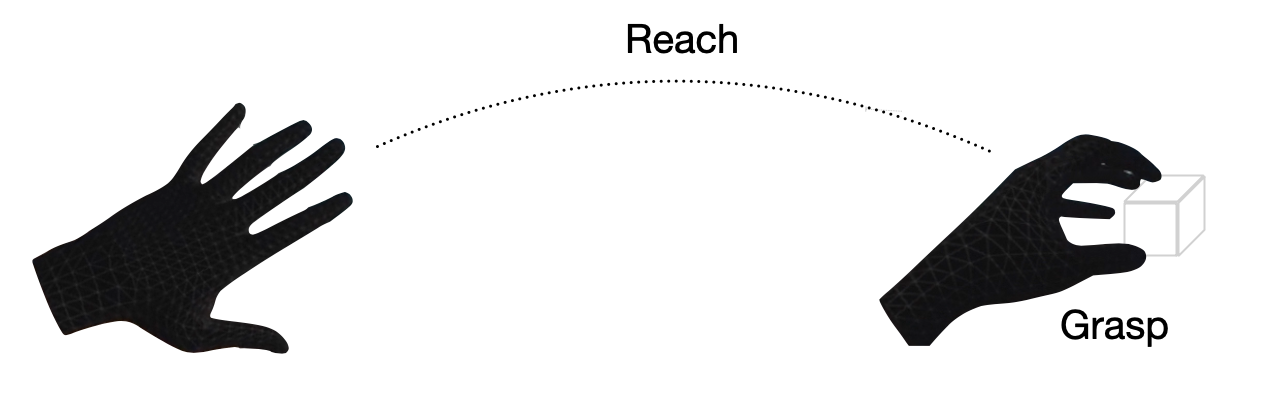}
    \caption{The process of "reach-to-grasp," beginning with reaching, where the hand is directed towards an object, and followed by grasping, where the hand conforms to the object's shape to secure it.}
    \label{fig:r2g}
\end{figure}

Reaching and grasping are two distinct yet interrelated actions in human-object interaction. As illustrated in Figure \ref{fig:r2g}. The process begins with reaching, where the hand is directed toward an object, followed by grasping, where the hand conforms to the objects shape to secure it. This reach-to-grasp motion has been extensively studied in various fields, including psychology, neuroscience \cite{10.1093/acprof:oso/9780195173154.001.0001, graspC}, and robotics \cite{doi:10.1177/027836499701600302, 8968559}, as it forms the foundation for most human-object interactions. 

Reaching involves transporting the hand to the vicinity of the target object, often characterized by smooth and efficient movements. The Minimum Jerk (MJ) model is commonly used to describe reaching trajectories, positing that human arms tend to follow paths that minimize the third derivative of position, or "jerk" \cite{flash1985coordination}. This model predicts that reaching motions are typically straight, with bell-shaped velocity profiles, and it has been used to model human reach trajectories in various settings. For example, Bratt et al. \cite{4398989} applied MJ model in a catching ball task to predict the human arm movement.

As the hand approaches the object, it begins to form the appropriate configuration for grasping. This anticipatory behavior, known as hand pre-shaping, is a critical aspect of the reach-to-grasp motion. Research by Chieffi et al. \cite{chieffi1993coordination} and Ansuini et al. \cite{ansuini2006effects} has shown that grasp formation is highly correlated with the shape and size of the target object, and this pre-shaping occurs well before the hand reaches the object. The hand adjusts its posture based on the object's properties and the intended action, allowing for a smooth transition from reaching to grasping.

The study of human grasping has been systematically organized into various taxonomies. Feix et al. \cite{7243327} defined a grasp as "every static hand posture with which an object can be held securely, irrespective of the hand orientation.", and classified 33 human hand configurations based on four key properties: virtual fingers, grasp types, opposition space, and thumb position.

The coordination between reach and grasp is essential for successful object manipulation. Although reaching and grasping can be modeled independently, studies have shown that these two processes are closely linked. For instance, Santello and Soechting \cite{santello1998gradual} demonstrated that hand shaping begins during the reaching phase and is influenced by both the object's properties and the intended action. This coordination ensures that the hand is properly configured by the time it reaches the object, enabling efficient and effective manipulation.

In VR, modeling the reach-to-grasp motion is important for creating realistic and responsive interactions. However, traditional models, such as MJ, often require most part of one motion sequence to be recorded before predictions can be made, which limits their applicability in real-time systems \cite{landi2019prediction}. To address this, machine learning techniques have been introduced to predict reaching and grasping behaviors more effectively, leveraging data-driven approaches to capture the complexity of human motion .

\subsection{Machine Learning for Predicting User Intentions in VR Interactions}
Machine learning has emerged as a powerful tool for predicting user intentions in VR interactions, especially in the context of hand tracking and gesture recognition. The high speed and dexterity of the human hand make precise tracking a challenging task, and traditional rule-based or optimization-based methods often fall short in capturing the complexity of human motion and mostly cannot be performed in real-time \cite{daiber2012towards, santello1998gradual, 9535175}. Machine learning, particularly neural networks, has shown great promise in overcoming these limitations by learning directly from large datasets of hand motion.

Recurrent Neural Networks (RNNs), including Long Short-Term Memory (LSTM) networks, are particularly well-suited for predicting sequential data, such as hand movements during reach-to-grasp actions. These models excel at capturing temporal dependencies in data, allowing them to predict the user's next action based on the sequence of past movements \cite{ghosh2017learning}. For instance, the work of Valkov et al. \cite{9535175} demonstrated the use of grasp features and LSTM to predict which object a user will interact with in real-time. Similarly, Aldrich et al. \cite{9417717} utilized LSTM to predict reach targets in a VR setting, still focusing on between-object predictions. While these models can perform in real-time, they still do not address the challenge of within-object prediction, where understanding the specific grasp location is crucial for more nuanced and accurate user intention prediction.

In another approach, by combining jerk-based trajectory fitting with online adaptation through neural networks, high accuracy in predicting movement targets was demonstrated in controlled environments, as shown by Landi et al. \cite{landi2019prediction}. However, this approach lacks the randomness and complexity inherent in real-world VR interactions.

While machine learning has made significant strides in improving user intention prediction, challenges remain. One key issue is the variability in user behavior. As demonstrated in the work of Vatavu et al., user-specific metrics can dramatically improve prediction accuracy \cite{vatavu2013automatic}, but this raises questions about the generalizability of models across different users. Additionally, the complexity of hand motion often requires models to balance between accuracy and computational efficiency, particularly in real-time applications where latency must be minimized.

In summary, machine learning offers a powerful framework for predicting user intentions in VR. However, the success of these models depends on their ability to handle the inherent variability and complexity of human motion. Future research should continue to explore ways to improve the generalization and adaptability of machine learning models in VR, ensuring that they can provide accurate, real-time predictions for a wide range of users and interactions.

\section{Data Collection}\label{chp:stuff}

\subsection{Data Sources}
The data were initially collected for an experiment to validate a geometrical method to predict users intention. 7 participants were recruited (three female), with ages ranging from 25 to 37 years (mean $=30$, std $=4$). All participants were right-handed. Participants used an Oculus Quest head-mounted display (HMD) without any additional sensors, controllers, or wearable devices. 



\begin{figure}[h]
    \centering
    \includegraphics[width=0.8\linewidth]{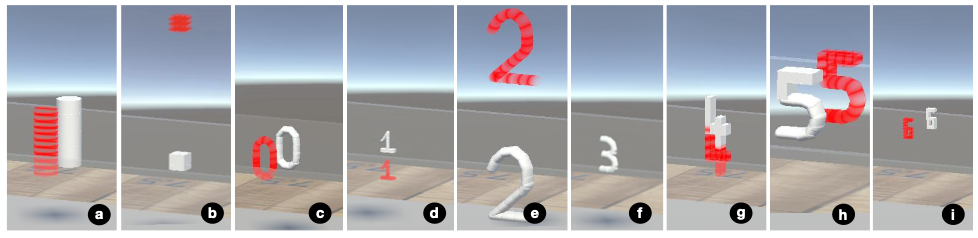}
    \caption{The task involves moving the white virtual object of varying shapes and sizes into the red phantom target area using a specified manipulation (e.g., hold, push, pull): (a) Medium Cylinder to be pulled; (b) Small Cube to be lifted; (c) Medium 0 to be pulled; (d) Small 1 to be pushed downward; (e) Large 2 to be lifted; (f) Small 3 to be touched; (g) Medium 4 to be pushed downward; (h) Large 5 to be pushed; (i) Small 6 to be pulled.}
    \label{fig:tasks}
\end{figure}

The 3D environment was created in Unity3D and compiled as an Android application. The scene included avatar hands 
provided by the Oculus Quest, as well as a virtual object to be manipulated (white) and its target location (red), as illustrated in Figure \ref{fig:tasks}. The virtual objects were not affected by Unity3D's physics engine (gravity) and were attached to the avatar's hands when a collision occurred, allowing them to move accordingly.


\subsection{Methods}
\subsubsection{Object Factors and Task Manipulation}
We controlled three factors for the experiment: objects size, objects, and the type of tasks. 

\begin{itemize}
    \item Size: 5 cm (small), 10 cm (medium), and 25 cm (large). These sizes were selected to represent objects relative to an average hand size (approximately 18 cm).
    \item Shape: Basic: Cube and Cylinder (Figure \ref{fig:tasks} (a,b)). Complex: digits 0 to 6 (Figure \ref{fig:tasks} (c-i)). Complex shapes provide multiple grasp opportunities. For example, the digit '1' could be grasped by its rectangular base, cylindrical trunk, or spherical top.
    \item Task: Hold, Pull, Push, Raise, and Push down (Figure \ref{fig:tasks}). Each task required different grasp types, opposition spaces, and thumb positions, reflecting common everyday manipulations.
\end{itemize}





\subsubsection{Data Recording}
Each participant was tested under 135 different conditions, corresponding to 5 manipulations × 9 shapes × 3 sizes, presented in a randomized order. The total number of trials was 945 (7 participants × 135 conditions).
We collected detailed data on each of the Oculus avatar hands’ phalanges and the corresponding objects of interest, including positions and orientations. All data were recorded at a frequency of 60 FPS.

\subsection{Data Analysis}
During the data collection process, we encountered an issue where data from User 2 was missing. As a result, we were left with 810 trials for analysis.

\begin{figure}[h]
    \centering
\includegraphics[width=0.45\linewidth]{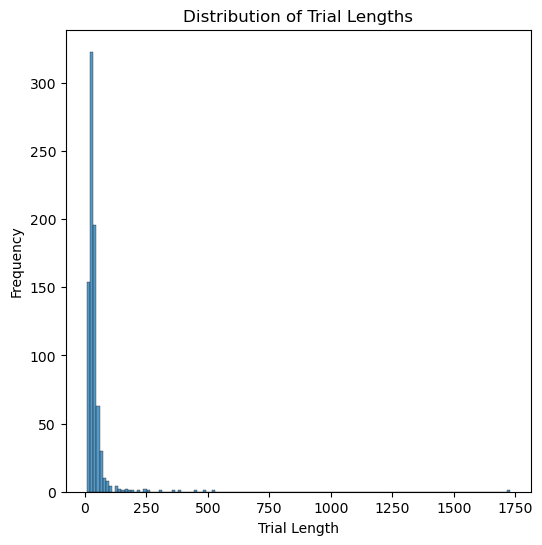}
    \caption{Distribution of trial lengths (in the number of frames).}
    \label{fig:length}
\end{figure}

On average, the trial length across all users was 41.69 frames ($\approx$ 4 seconds). The longest trial was 1728 frames, corresponding to configuration 106 (Raise, 3, Small) for User 4. The shortest trial lasted only 9 frames, corresponding to configuration 87 (Pull, Cube, Large), also for User 4. The majority of trials had lengths within the range of 0 to 100 frames. Figure \ref{fig:length} provides the distribution of trial lengths. The maximum moving distance of user's palm was 1.52 m, the minimum was 0.04 m, and the average was 0.46 m.

Upon further investigation, we found that, for all users, the first trial tended to be the longest. This is likely because users were still adapting to the virtual reality environment and familiarizing themselves with the tasks during their initial interactions.

\section{Users Intentions Prediction as a Classification Problem}

\subsection{Motivation}
Initially, we formalized the prediction of user intentions in virtual reality, particularly in the context of grasping, as a classification problem. This approach is naturally supported by the structured nature of our data, which includes distinct labels for object size, shape, and manipulation type. These labels provide clear categories that align with the goals of classification, making it an intuitive framework for predicting specific user actions. By leveraging these labels, we can apply supervised learning techniques to categorize each user interaction, thereby predicting their intentions with precision.

The use of classification is further reinforced by foundational research in grasp taxonomy. Grasping is inherently dependent on both the task and the object \cite{10.3389/fnhum.2012.00117, napier1956prehensile}, which means that different scenarios require different hand configurations. For instance, the way a user grasps a bottle varies depending on whether they intend to drink from it or pass it to someone else. These variations are effectively captured by classification methods, which allow us to model and predict user intentions based on specific features of the task and object.

\begin{figure}[h]
    \centering
    \includegraphics[width=\linewidth]{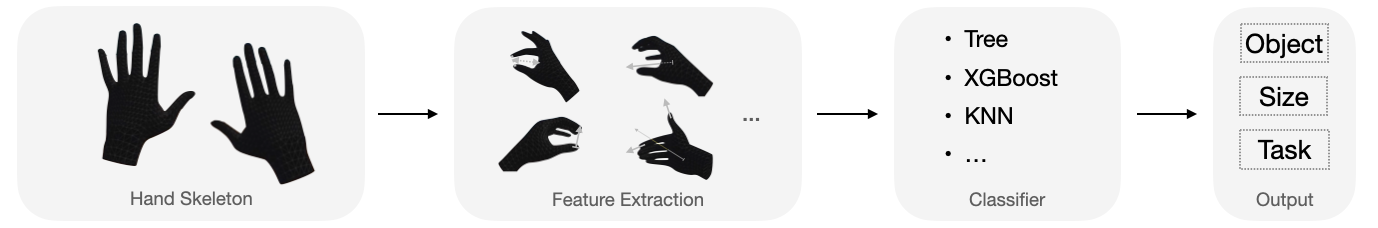}
    \caption{The process of predicting user intentions as a classification problem. Starting from hand skeleton data, grasp taxonomy-based features are extracted and fed into various classifiers (e.g., Linear models, XGBoost, KNN) to predict user actions.}
    \label{fig:classification}
\end{figure}

In addition to its theoretical foundation, classification is a classical and effective approach for handling grasp-related tasks, as demonstrated by prior research. Work by Iberall \cite{doi:10.1177/027836499701600302} and Cutkosky et al. \cite{Cutkosky1990} has shown that understanding the configuration of virtual fingers and the forces they apply can be systematically modeled through classification. In \cite{10.1145/3544549.3585773}, vectors between phalanges and undirected hand velocity are used as features for classification. Feature engineering, combined with classification algorithms, has proven to be a robust method for analyzing and predicting complex hand-object interactions, which is directly applicable to our objectives. Figure \ref{fig:classification} illustrates this process, which outlines the steps from hand skeleton data acquisition to the classification of user intentions.

Classifying user intentions also offers significant advantages in terms of data analysis. By categorizing user actions, we can gain deeper insights into the relationships between various features, such as object size, shape, and the type of manipulation involved. This structured analysis helps us not only to improve prediction accuracy but also to better understand the underlying dynamics of user interactions in virtual environments, which is crucial for refining our models and enhancing their performance.

Thus, the decision to approach user intention prediction as a classification problem is driven by the clear structure of our data, the support of existing research in grasp taxonomy, and the need to thoroughly understand the nuances of user behavior in virtual reality. This framework allows us to build on established methodologies while gaining valuable insights into the data, leading to more reliable and accurate predictions.

\subsection{Feature Extraction}
\begin{figure}[h]
    \centering
    \includegraphics[width=0.95\linewidth]{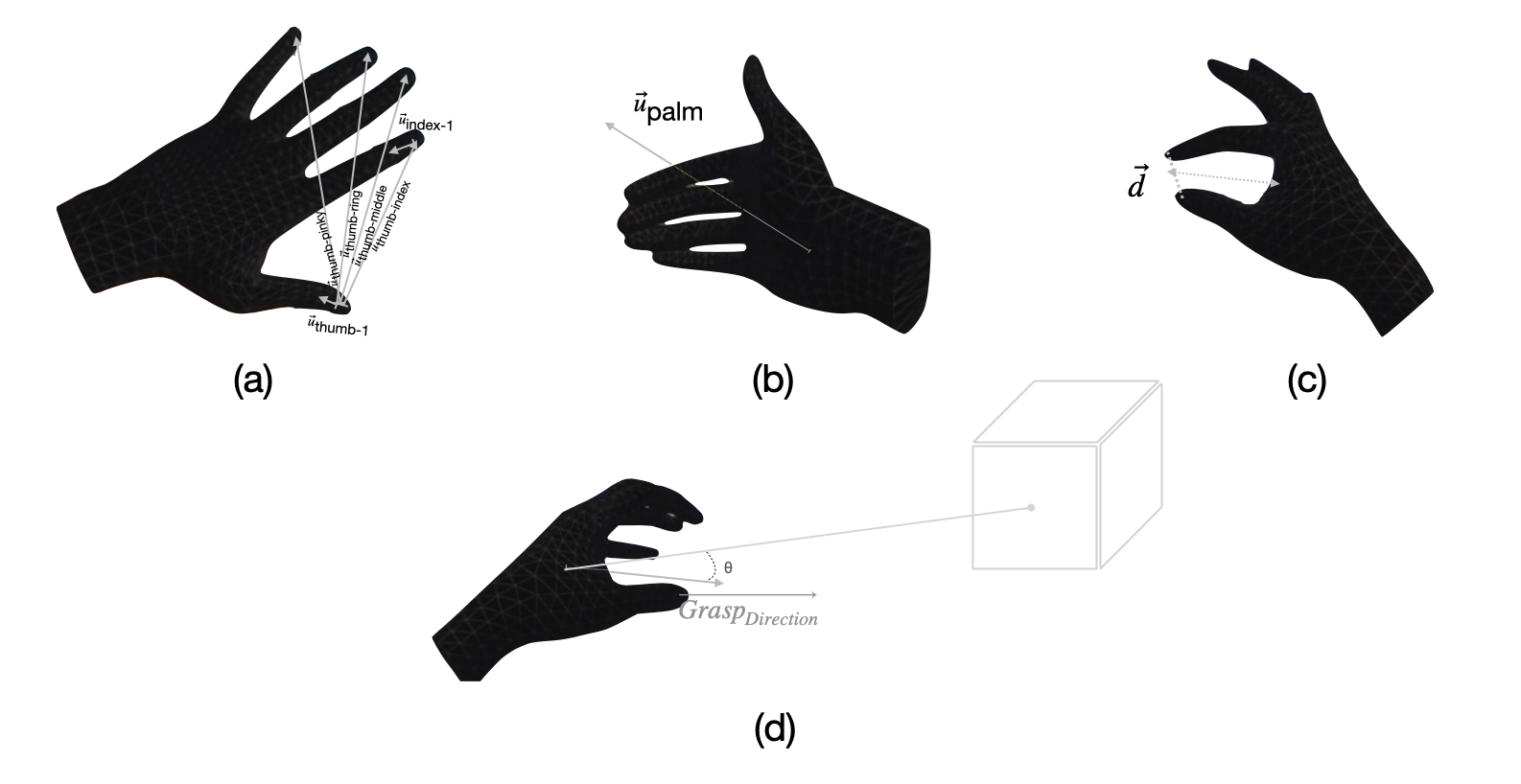}
    \caption{Illustration of features extracted for grasp prediction. (a) Thumb tip to finger tip vectors ( $\left.\vec{u}_{\text {thumb-index}}, \vec{u}_{\text {thumb-middle}}, \vec{u}_{\text {thumb-ring}}, \vec{u}_{\text {thumb-pinky}}\right)$ and thumb/index tip to proximal joint vectors $\left(\vec{u}_{\text {thumb-1}}, \vec{u}_{\text {index-1}}\right)$; (b) Palm vector approximation $\left(\vec{u}_{\text {palm}}\right)$; (c) Grasp depth vector $(\vec{d})$; (d) Palm-to-object angle.}
    \label{fig:features}
    \end{figure}

As noted in Valkov et al. \cite{10.1145/3544549.3585773}, vectors between adjacent phalanges can be used as features in predicting user actions. However, these vectors may appear identical across different tasks. For example, the finger configurations for push and pull might produce nearly identical vectors. Thus, through iterative trial and error, we identified a set of features that are more suitable for our task:

\begin{itemize}
    \item Thumb Tip to All Other Tips:
    Vectors from the thumb tip to the tips of the other fingers, denoted as $\vec{u}_{\text {thumb-index}}, \vec{u}_{\text {thumb-middle}}, \vec{u}_{\text {thumb-ring}}$ and $\vec{u}_{\text {thumb-pinky}}$ (Figure \ref{fig:features}(a)). These vectors, represented in 3D space, capture key aspects of the grasping pose. Notably, the thumb-to-index vector ($\vec{u}_{\text {thumb-index}}$) is widely recognized for describing grasp apertures, which is a commonly used approach for characterizing hand poses \cite{graspC}. The length of this vector also serves as an additional feature, capturing the aperture during grasping.

    \item Thumb and Index Tip to Proximal Joint Vectors: These vectors ($\vec{u}_{\text {thumb-1}}$ and $\vec{u}_{\text {index-1}}$) capture the degree of finger flexion (Figure \ref{fig:features}(a)), indicating how much the fingers are bent during grasping. By extracting the vectors from the thumb tip and index tip to their corresponding proximal joints, we gain valuable information about the hand's configuration during various tasks.

    \item Palm Vector Approximation: In environments where only a limited number of trackers are available (typically on the thumb and index finger), direct extraction of palm orientation is not feasible. To approximate the palm orientation, we derive a vector using the cross-product of the thumb-to-index vector, $\vec{u}_\text{palm}=\vec{u}_{\text {thumb-index }} \times \vec{z}_i$, where $\vec{z}_i$ represents the local Z axis of the index finger. This approximates the palm's orientation using available data (Figure \ref{fig:features}(b)).

    \item Grasp Depth: Geometrically, a grasp occurs within the volume defined by the opposition vector and the palm. Even in cases where the hand is not fully closed (e.g., non-prehensile grasp), this theory remains valid, as the contact is still aligned with the opposition vector. Since grasp aperture and finger grip are formed during the motion, we extract the object's "in-depth" contact point early in the grasp. We define the grasp depth, $\vec{d}$, as the vector between the palm center and the midpoint of the grasp aperture (Figure \ref{fig:features}(c)):

    $$
    \vec{d}=\overrightarrow{P_c P_m}
    $$

    where $P_m$ is the midpoint between the thumb and index tips and $P_c$ represents the palm center. The length of grasp depth is also served as a feature.

    \item Palm-to-Object Angle: The angle between the direction of palm movement and the center of the object serves as a feature to indicate the user’s intended direction of interaction with the object (Figure \ref{fig:features}(d)). This angle provides context on how the user plans to approach or manipulate the object.
    
\end{itemize}

In total, we extracted 11 features, all vectors are in 3D, and considered both hands in our analysis.

\subsection{Experimental Results}
To evaluate our approach, we first selected fixed-length time sequences leading up to the grasp event. We hypothesized that the closer the time window is to the contact moment, the more informative the features would be for predicting the corresponding labels. This is because earlier frames may contain significant noise, such as the user adjusting their hand positioning.

We applied a 5-fold cross validation with a Random Forest classifier, achieving the following average accuracies in the time window (-1 to -5): Object Accuracy: 0.96, Task Accuracy: 0.94, Scale Accuracy: 0.96, and Overall Accuracy: 0.90.

As illustrated in Figure X, classification accuracy increases as the model approaches the moment of grasp. The highest accuracies are observed in the last few time windows (-1 to -5), which aligns with our expectation that features closer to the grasp event carry more predictive information.

However, it is important to note that only trials that are long enough are included in the earlier time windows. The vertical dashed line at the time window (-6 to -10) marks the cutoff where all 810 trials (810 $\times$ 5 = 4050 data points) are included. Beyond this point, the increase in accuracy can be partially attributed to the reduced data volume, as shorter trials do not span the earlier windows, simplifying the task and reducing variability.

\begin{figure}[h]
    \centering
    \includegraphics[width=\linewidth]{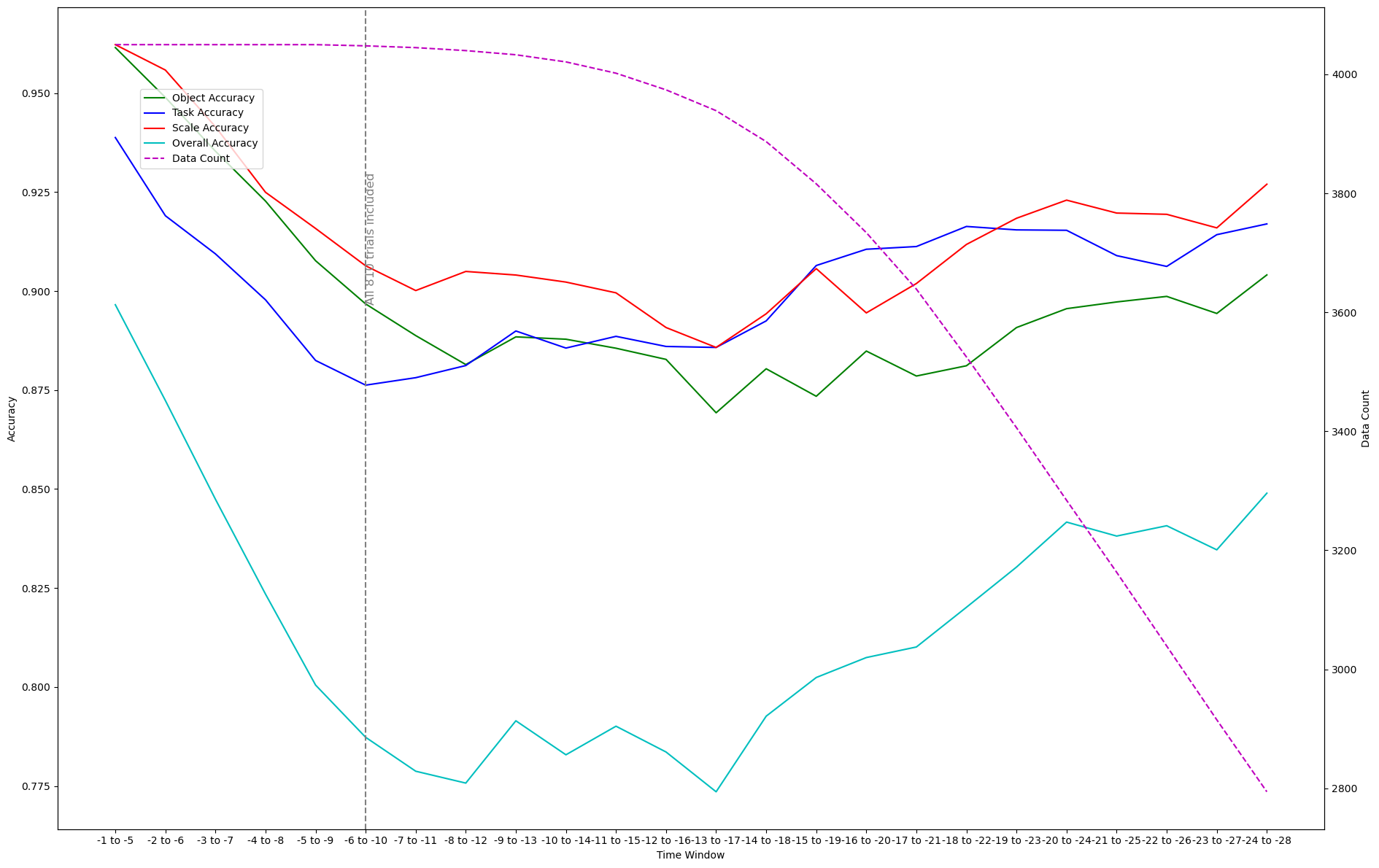}
    \caption{Classification accuracy over different time windows given by random forest classifier, where 0 represents the moment of grasp. Accuracy increases as the window approaches the grasp event, with the highest accuracies observed in the last time window (-1 to -5). The dashed vertical line indicates the cutoff point (-6 to -10) where all 810 trials are included. Beyond this point, accuracy continues to rise due to a decrease in data volume, as shorter trials do not span earlier time windows, making the task less complex.}
    \label{fig:count_acc}
\end{figure}

We compared the performance of several classifiers using 5-fold cross-validation to determine the best model for our task. Among the classifiers tested, the Random Forest model performed the best, achieving the highest average accuracy across all categories. The detailed results for each model are shown in Table \ref{tab:5fold_results}. We chose not to use deep learning due to the relatively small dataset (810 trials), which may not be sufficient for training deep networks effectively. Additionally, traditional models like Random Forests offer better interpretability and are more computationally efficient for our current setup.

\begin{table}[h!]
\centering
\caption{Average accuracy across different classifiers using 5-fold cross validation.}
\begin{tabular}{|l|c|c|c|c|}
\hline
\textbf{Classifier}    & \textbf{Object} & \textbf{Size} & \textbf{Task} & \textbf{Overall} \\ \hline
\textbf{Decision Tree} & 0.79                     & 0.86                   & 0.73                    & 0.54                      \\ \hline
\textbf{Random Forest} & 0.96                     & 0.96                   & 0.94                    & 0.90                      \\ \hline
\textbf{XGBoost}       & 0.95                     & 0.96                   & 0.94                    & 0.88                      \\ \hline
\textbf{KNN (k=1)}     & 0.93                     & 0.95                   & 0.91                    & 0.88                      \\ \hline
\end{tabular}
\label{tab:5fold_results}
\end{table}

As shown in Table \ref{tab:5fold_results}, the Random Forest model outperformed the other classifiers, with an overall accuracy of 0.90, followed by XGBoost and KNN. The Decision Tree model, in contrast, had the lowest overall accuracy of 0.54, highlighting the importance of ensemble methods and the potential benefits of using multiple trees over a single decision tree.

To evaluate whether the method could generalize to unseen users, we also conducted a leave-one-user-out test. This approach tests the model's ability to predict user intentions when trained on all users except one, which is left out for validation. However, the results from this evaluation were significantly worse, indicating that our current method struggles to generalize to new users. These findings suggest that the model heavily relies on user-specific patterns and is not yet robust enough to handle unseen individuals. The performance under the leave-one-user-out cross validation setting is summarized in Table \ref{tab:leave_one_out_results}.

\begin{table}[h!]
\centering
\caption{Average accuracy across different classifiers using leave-one-user-out cross validation.}
\begin{tabular}{|l|c|c|c|c|}
\hline
\textbf{Classifier}    & \textbf{Object} & \textbf{Size} & \textbf{Task} & \textbf{Overall} \\ \hline
\textbf{Decision Tree} & 0.40                       & 0.55                     & 0.26                      & 0.06                        \\ \hline
\textbf{Random Forest} & 0.51                       & 0.64                    & 0.31                      & 0.12                        \\ \hline
\textbf{XGBoost}       & 0.51                       & 0.68                     & 0.30                      & 0.11                        \\ \hline
\textbf{KNN (k=1)}     & 0.32                       & 0.52                     & 0.25                      & 0.05                        \\ \hline
\end{tabular}
\label{tab:leave_one_out_results}
\end{table}

The poor performance in this setting underscores the need for further analysis into improving the generalization capabilities of our model.

\subsection{Error Analysis}

\subsubsection{Confusion Matrices}

\begin{figure}[h]
    \centering
    \includegraphics[width=\linewidth]{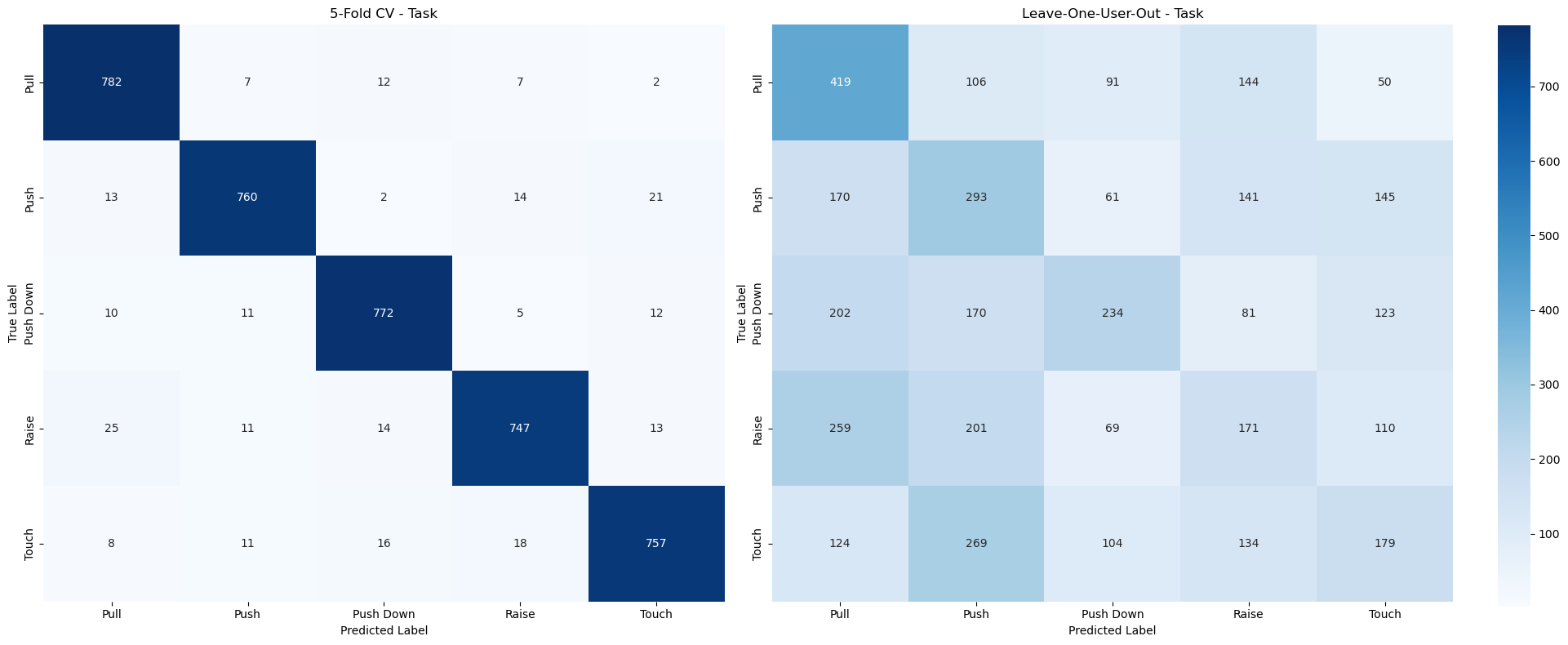}
    \caption{Confusion matrices for task classification: (left) 5-fold cross validation, (right) Leave-one-user-out cross validation.}
    \label{fig:conftask}
\end{figure}

\begin{figure}[h]
    \centering
    \includegraphics[width=\linewidth]{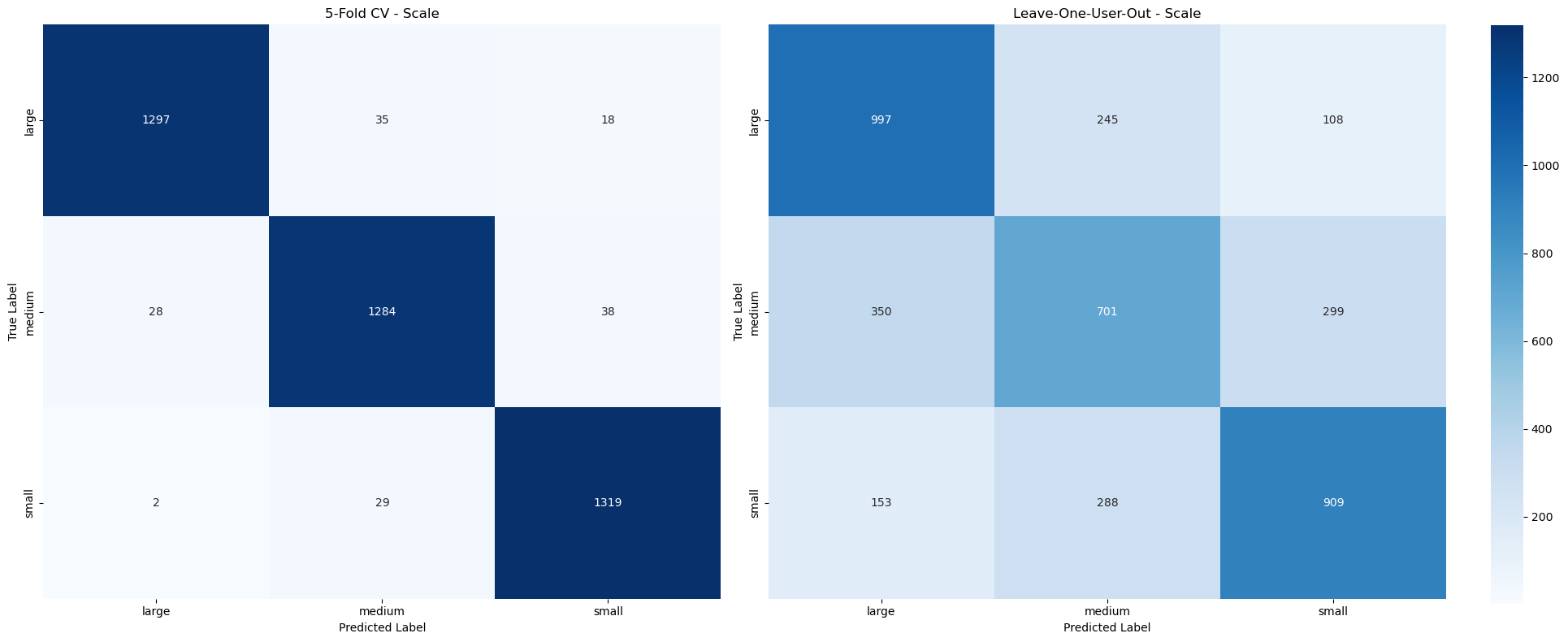}
    \caption{Confusion matrices for size classification: (left) 5-fold cross validation, (right) Leave-one-user-out cross validation.}
    \label{fig:confscale}
\end{figure}

To gain deeper insights into the challenges faced by our classifiers, we analyzed the confusion matrices for the 5-fold cross validation and leave-one-user-out cross validation approaches shown in Figure \ref{fig:confobj}, \ref{fig:conftask} and \ref{fig:confscale}. The results revealed several key patterns:
\begin{figure}
    \centering
    \includegraphics[width=\linewidth]{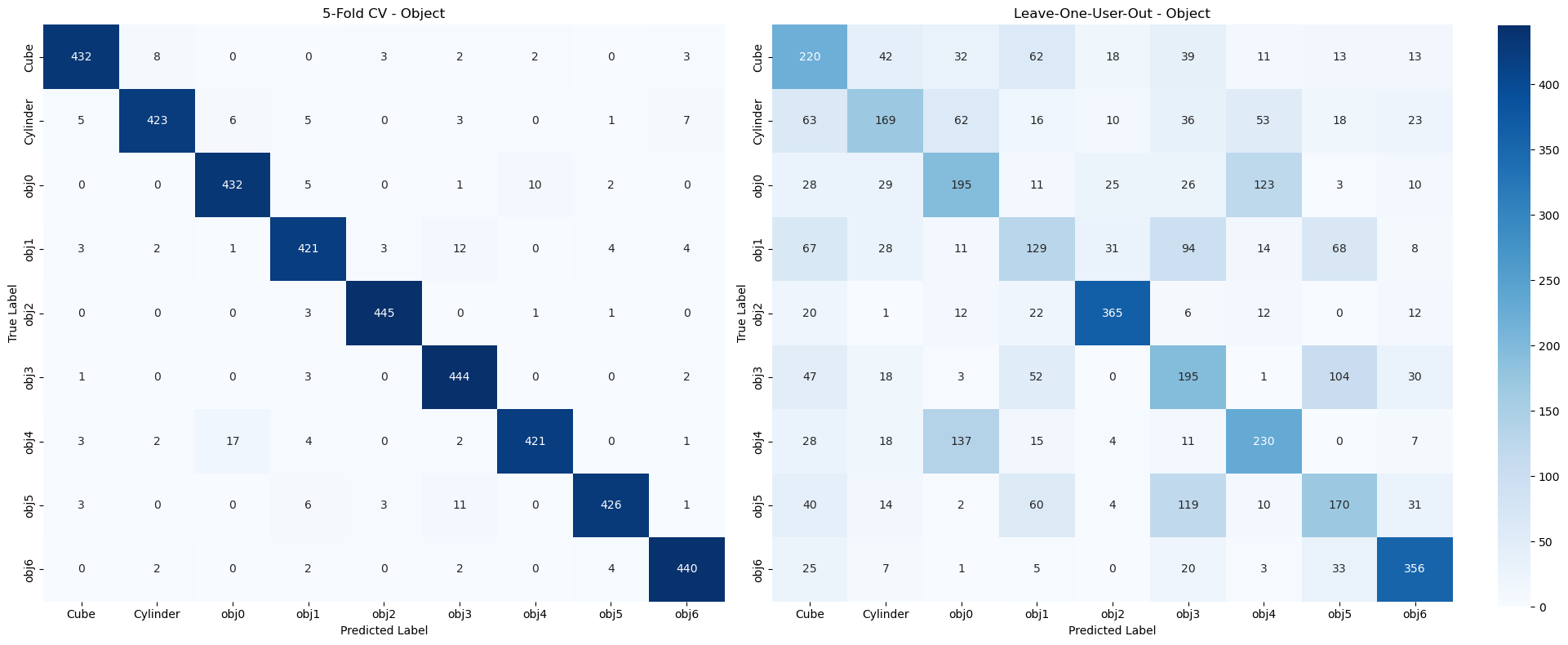}
    \caption{Confusion matrices for object classification: (left) 5-fold cross validation, (right) Leave-one-user-out cross validation.}
    \label{fig:confobj}
\end{figure}
\FloatBarrier
\FloatBarrier
\begin{itemize}
    \item Error Magnification in Leave-One-User-Out: Categories that were difficult to classify in the 5-fold CV scenario became even more error-prone in the leave-one-user-out setting. 
    
    \item Task Classification Confusion (Figure \ref{fig:conftask}): The confusion matrix for tasks in the leave-one-user-out setting shows a significant level of confusion across all tasks. This suggests that the model struggles to distinguish between different tasks, which may be due to the variability in how different users execute the same task. 
    \item Scale Classification Patterns (Figure \ref{fig:confscale}): The scale classification results were intuitive. The confusion was more concentrated between sizes that are close in value, such as small and medium.
\end{itemize}

These observations highlight the inherent challenges of using classification models in the context of user intention prediction, especially when faced with high variability in user behavior. This leads us to further qualitative analysis to better understand where the classifiers are failing.

\subsubsection{Qualitative Analysis}
In this section, we analyze the reasons behind the lower performance of our classifiers using leave-one-user-out cross validation, particularly the consistently lower accuracy in predicting tasks compared to object and scale labels. A key issue stems from the approach of labeling user intentions, which proves to be inherently flawed. Users have the freedom to perform tasks in varied ways, leading to significant variability in how tasks are executed.

\begin{figure}[h]
    \centering
    \begin{minipage}[b]{0.3\textwidth}
        \centering
        \includegraphics[width=\textwidth, height=5cm]{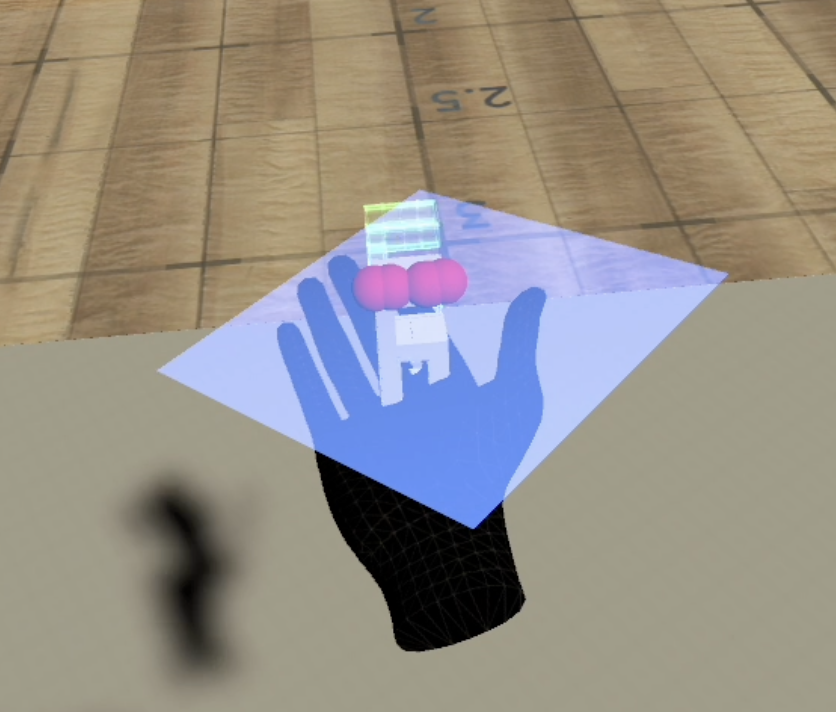}
        \caption*{(a)}
    \end{minipage}
    \hfill
    \begin{minipage}[b]{0.3\textwidth}
        \centering
        \includegraphics[width=\textwidth, height=5cm]{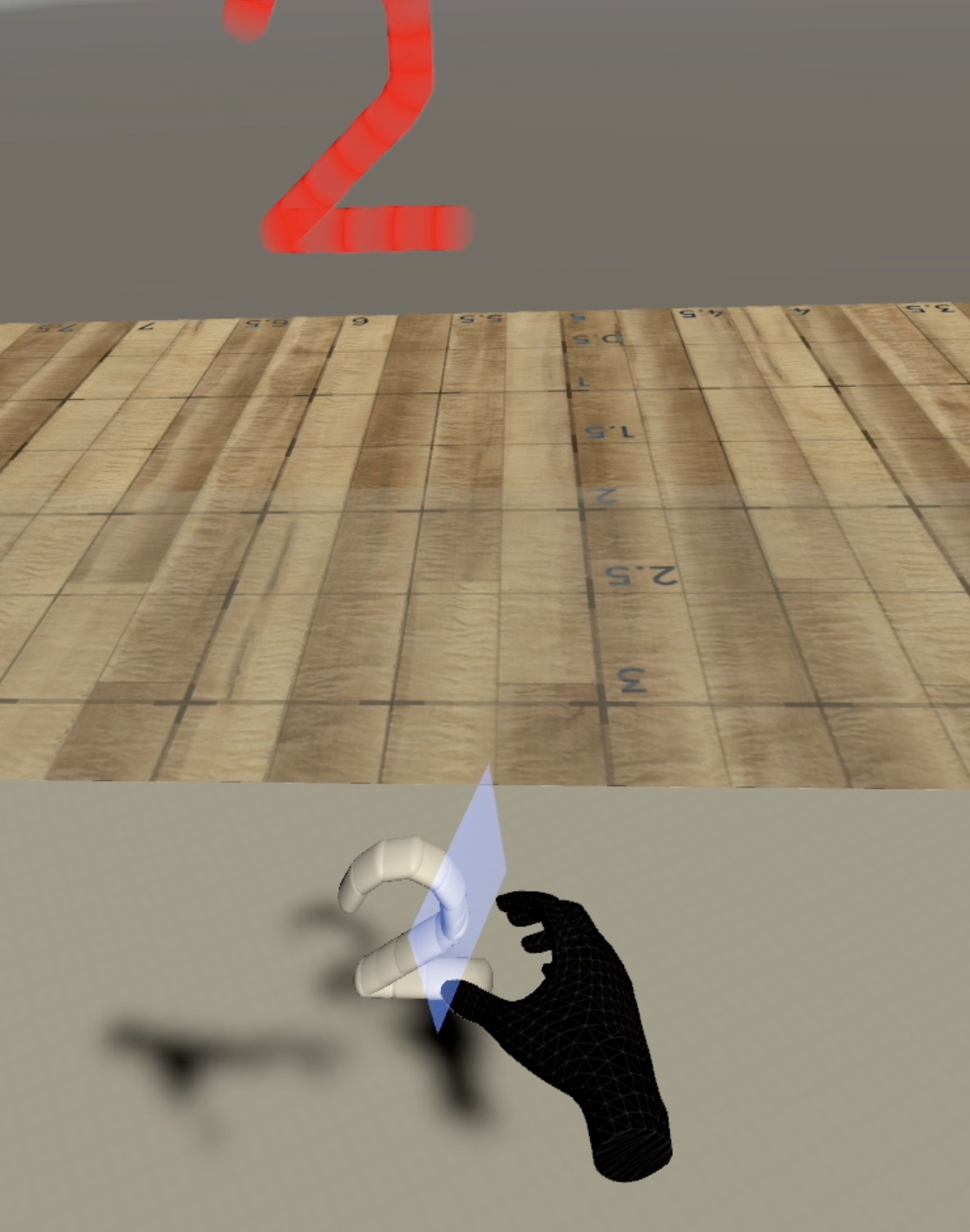}
        \caption*{(b)}
    \end{minipage}
    \hfill
    \begin{minipage}[b]{0.3\textwidth}
        \centering
        \includegraphics[width=\textwidth, height=5cm]{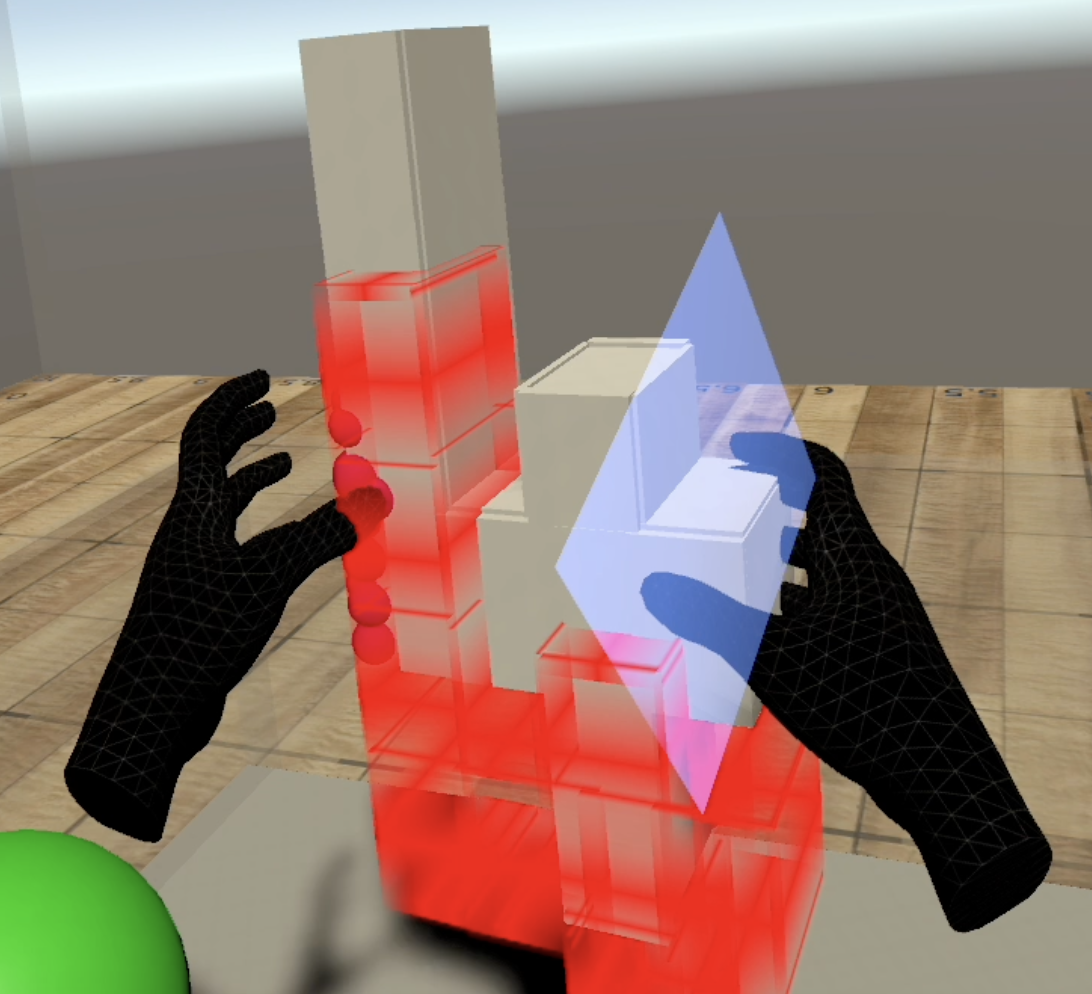}
        \caption*{(c)}
    \end{minipage}
    \caption{Examples of common misclassifications: (a) A "Touch" action being misclassified as "Raise", (b) A "Raise" action misclassified as "Pull", and (c) A "Push down" action misclassified as "Pull".}
    \label{fig:misclassifications}
\end{figure}

For instance, there are several scenarios where the classifiers almost consistently make mistakes. One common error is misclassifying a "Touch" action, where a user touched the bottom of an object, as a "Raise" action (see Figure \ref{fig:misclassifications}(a)). In another example, a user who grasped an object from the side with the intention of "Raise" ended up having their action misclassified as a "Pull" (see Figure \ref{fig:misclassifications}(b)). Additionally, in a cases where an user was instructed to "Push down," the classifiers occasionally misinterpreted the action as a "Pull," possibly because the user's perspective led them to unconsciously perform a pulling motion (see Figure \ref{fig:misclassifications}(c)). This shows how perspective and subtle differences in execution can significantly impact classification.

\begin{figure}[h]
    \centering
    \begin{minipage}[b]{0.45\textwidth}
        \centering
        \includegraphics[width=0.5\textwidth]{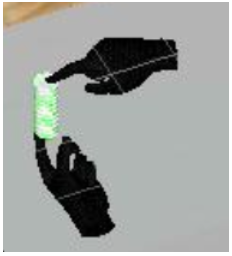}
        \caption*{(a)}
    \end{minipage}
    \hfill
    \begin{minipage}[b]{0.45\textwidth}
        \centering
        \includegraphics[width=0.5\textwidth]{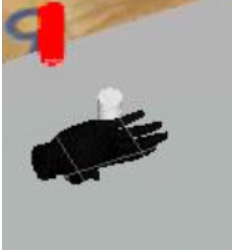}
        \caption*{(b)}
    \end{minipage}
    \caption{Examples of variable user behavior: (a) "Touch" by holding the object between two fingers, (b) "Push" action performed with an upside-down hand position.}
    \label{fig:user_variability}
\end{figure}

Moreover, the variability in user behavior further complicates the task of accurate classification. For example, an user tried to "Touch" an object by holding it between two fingers (see Figure \ref{fig:user_variability}(a)), and an user might "Push" an object with an upside-down hand position (see Figure \ref{fig:user_variability}(b)). Such variations introduce noise into the data, making it difficult for the classifiers to generalize across different users and their unique approaches to tasks.

\begin{figure}[h]
    \centering
    \begin{minipage}[b]{0.45\textwidth}
        \centering
        \includegraphics[width=\textwidth, height=7.2cm]{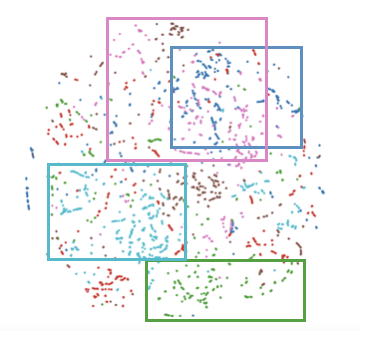}
        \caption*{(a)}
    \end{minipage}
    \hfill
    \begin{minipage}[b]{0.45\textwidth}
        \centering
        \includegraphics[width=0.9\textwidth, height=6.8cm]{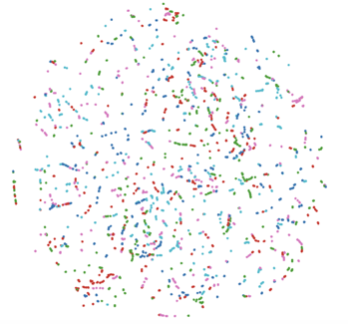}
        \caption*{(b)}
    \end{minipage}
    \caption{t-SNE visualization of feature clusters: (a) Clusters annotated by user, showing distinct separation, (b) Clusters annotated by task, with less distinct separation.}
    \label{fig:tsne_clusters}
\end{figure}

To better understand these challenges, we conducted a t-SNE analysis, a dimensionality reduction technique used to visualize high-dimensional data in two dimensions. The resulting clusters from our feature extraction show that when annotated by user, the data forms more distinct clusters, which are highlighted with corresponding colored borders (see Figure \ref{fig:tsne_clusters}(a)). In contrast, when annotated by task, the clusters are less defined and more dispersed (see Figure \ref{fig:tsne_clusters}(b). This finding suggests that the primary variation in the data is user-driven rather than task-driven, further complicating the task classification process.

In conclusion, our analysis indicates that using classification to predict user intentions may not be a viable approach, as the variability in user behavior and the flexible nature of task execution introduce significant challenges. A different strategy may be required to effectively model and predict user intentions in virtual reality environments.

\subsection{Discussion}
The analysis of our experimental results and error analysis reveals several key findings that contribute to our understanding of the challenges involved in predicting user intentions in VR environments using classification-based methods. These insights provide direction for future research and potential improvements in this area.

\begin{enumerate}
    \item Impact of User Variability: One of the most significant findings is the impact of user variability on classification performance. Different users exhibit distinct behaviors when performing the same tasks, which leads to inconsistencies in how the model interprets and predicts their actions. This is especially evident in task classification, where user-specific variations in task execution resulted in a high degree of confusion.

    This suggests that user intention prediction models in VR must account for the high variability in human behavior. Approaches that are heavily dependent on user-specific data may not generalize well to new users. Future models may need to incorporate user-adaptive mechanisms that can dynamically adjust to individual behaviors or consider training models on a broader dataset that captures a wider range of user interactions.

    \item Task Complexity and Ambiguity: Our analysis indicates that tasks with similar hand configurations (e.g., "Push" and "Pull") often result in misclassifications. This is partly due to the inherent ambiguity in how tasks are labeled and performed. Even subtle differences in user intent or perspective can lead to drastically different outcomes in classification, as seen in our qualitative analysis.

    This highlights the need for more robust task definitions and possibly a reconsideration of how tasks are labeled and modeled. A potential solution could be to incorporate temporal dynamics into the model, where sequences of actions, rather than isolated frames, are used to determine the task. This would allow the model to capture the progression of an action and provide more accurate predictions.

    \item Limitations of Classification-Based Approaches: Our results suggest that classification-based approaches may not be well-suited for predicting user intentions in complex, real-time VR interactions. While classification offers a structured way to categorize actions, it may oversimplify the nuances of human behavior, especially in a dynamic environment like VR. The findings from the t-SNE analysis support this, showing that user behavior is more distinctly clustered by individual differences rather than task-related features.

    This calls into question whether classification is the most effective framework for user intention prediction in VR. Future research should explore alternative approaches, such as regression models that predict continuous outcomes, or hybrid models that combine classification with other methods to capture the complexity of user interactions.
\end{enumerate}

In conclusion, while classification provides a structured approach to predicting user intentions, it is limited by the variability and complexity of human behavior in VR environments. Addressing these challenges will require more flexible and adaptive approaches that go beyond traditional classification methods.

\section{Users Intentions Prediction as a Regression Problem}
\subsection{Motivation}
Following our analysis of classification models, we turn to regression-based approaches to overcome the limitations of mislabeling and the variability in user actions. Unlike classification, which forces discrete labels on dynamic interactions, regression allows for continuous predictions, making it better suited for capturing the varied nature of user behavior.

Regression models enable us to predict both the posture and position of the hand over time, directly addressing the key questions of \textit{when}, \textit{where}, and \textit{how} a grasp will occur. This is particularly important for haptic systems, as precise predictions of the hand’s posture during a grasp allow for the delivery of tailored haptic responses, enhancing the realism of the VR experience. This approach also supports within-object predictions, where understanding the specific interactions with different parts of an object is crucial for creating immersive and accurate feedback.

An additional strength of regression models is their capacity to incorporate biomechanical principles, such as the Minimum Jerk Model, which describes smooth and natural human motion trajectories. By integrating this model, we can predict not only the final grasp position but also the timing of the hand’s movement as it approaches and interacts with an object, leading to more realistic predictions.

We decompose the task of predicting user intentions into two subproblems:

\begin{enumerate}
    \item Predicting the Position and Time When Grasp Happens: This involves predicting the hand’s final position and the exact moment when the grasp occurs, addressing both \textit{when} and \textit{where} the grasp will take place.
    \item Predicting the Posture of the Hand When Grasp Happens: This focuses on predicting the specific configuration of the fingers and hand during the grasp, addressing the \textit{how} aspect of the interaction.
\end{enumerate}

By framing user intention prediction as a regression problem, we aim to surpass the limitations of classification models and provide more nuanced and adaptable predictions that account for the complexity of real-world VR interactions. This not only enhances predictive accuracy but also facilitates more responsive and immersive haptic feedback in virtual environments.

\subsection{Predicting the Position and Time When Grasp Happens}
\subsubsection{Problem Setup}
\textbf{Input Data:}

Since all users in our study are right-handed, we focus exclusively on the right-hand data. The primary input to our model is the 3D position of the palm as it moves towards an object during a grasp. To determine the optimal time for prediction, we focus on data from the last two seconds leading up to the grasp. We use this 2-second time window because most grasping actions in our task occur within this duration. Our assumption is that features become more predictive as the hand approaches the object.

The input data is structured as a time-series of 3D coordinates for the palm's position at each time frame, denoted as $\mathbf{P}(t)=[x(t), y(t), z(t)]$. Additionally, we incorporate the time differences between consecutive frames, denoted as $\Delta t_i=t_{i}-t_{i-1}$, to account for the uneven sampling rate of the tracking device. Thus, each input frame includes both the palm's 3D position $\mathbf{P}\left(t_i\right)$ and the time difference $\Delta t_i$ between the frames.

To explore the most effective point in the trajectory for prediction, we apply a sliding window technique. For each sequence, we generate multiple data points by considering different segments of the movement. For example, one data point might include movement data from -2 to -0.5 second before the grasp, while another might include data from -2 to -1 seconds before the grasp. The prediction targets for these two data points will have the same palm position but different predicted times to reach the grasp event. This approach allows us to evaluate the model's performance at various points in the movement trajectory.

Formally, the input to the model for a sequence of $n$ time steps can be represented as:

$$
\mathbf{X}=\left\{\left[\mathbf{P}\left(t_1\right), \Delta t_1\right],\left[\mathbf{P}\left(t_2\right), \Delta t_2\right], \ldots,\left[\mathbf{P}\left(t_n\right), \Delta t_n\right]\right\}
$$

\textbf{Prediction Targets:}

The output of the model includes two components:

$$
\mathbf{Y}=\left[\mathbf{P}_{\text {grasp }}, T_{\text {grasp }}\right]
$$

\begin{enumerate}
    \item Predicted Palm Position: The 3D coordinates of the palm at the moment of grasp, denoted as $\mathbf{P}_{\text {grasp }}=\left[x_{\text {grasp }}, y_{\text {grasp }}, z_{\text {grasp }}\right]$. This provides the final position of the palm at the moment of contact with the object.
    \item Predicted Time to Grasp: The remaining time until the grasp event occurs, denoted as $T_{\text {grasp }}$. This indicates how much time is left before the user starts the grasp, which is critical for aligning haptic feedback with the user's action in real-time.
\end{enumerate}

\subsubsection{Method}
\begin{enumerate}
    \item \textbf{Minimum Jerk:} The Minimum Jerk Trajectory (MJT) is a well-established model for human motion, which postulates that natural hand movements minimize the rate of change of acceleration (jerk). This results in smooth and efficient motion that is characteristic of many human tasks. The jerk cost function is defined as:
    $$
    M=\int_0^{t_f}\|\dddot {x}(t)\|^2 d t
    $$
    where $t_f$ is the duration of the movement. When the movement takes place along a straightline segment, the position of the hand at time $t$ can be described by the following equation:
    $$
    x(t)=x_0+\left(x_f-x_0\right)\left(6 \tau^5-15 \tau^4+10 \tau^3\right)
    $$
    where $\tau= \frac{t}{t_f}$, the normalized time, varies from 0 to 1. $x_0$ and $x_f$ represent the initial and final positions, respectively. The velocity profile of the MJT can be derived as:
    
    $$
    v(t)=\dot{x}(t)=x_f\left(30 \tau^4-60 \tau^3+30 \tau^2\right)
    $$
    
    In our task, we aim to predict the final position of the hand $\mathbf{P}_{\text{grasp}} = x_f$ and the time to reach the object $T_{\text{grasp}} = t_f$. 
    
    Using the Minimum Jerk model for human motion prediction involves detecting the start of the movement, which we define as when the hand's velocity exceeds 3 cm/s. To calculate hand velocity while minimizing noise, we applied a smoothing differentiation filter using the Savitzky-Golay algorithm \cite{LUO2005122}. We then fit the MJT parameters (final position $x_f$ and time $t_f$) using Scipy's 'curve\_fit' function, which applies a least-squares optimization. The bounds for the final position are set to $\pm 0.5$ meters around the last known position, and the bounds for the remaining movement duration are set to 0-2 seconds, as most grasping tasks are completed within this timeframe.
    \item \textbf{Machine Learning:} Early experiments using the Minimum Jerk Trajectory (MJT) for predicting the final grasp position and time revealed limitations in fitting accuracy, even when utilizing the full motion sequence. This led us to explore more advanced approaches. Some research proposes using neural networks to predict additional points in the trajectory before applying MJT fitting \cite{8968559}. However, in our context, this approach proved suboptimal. Consequently, we chose to directly apply neural networks, specifically LSTM networks, to the time-series data to predict the final grasp position and time. Additionally, we explored a hybrid model that combines the output of the MJT with the LSTM model to benefit from both biomechanical principles and data-driven learning. Below, we detail both approaches:

    \begin{itemize}
    \item \textbf{LSTM:} The pure LSTM model processes the time-series data, which consists of 3D palm coordinates and time differences. The LSTM layer has 64 units with a dropout rate of 0.2 to prevent overfitting. After passing through the LSTM layer, the output is processed by a fully connected layer with 16 units and a ReLU activation function to generate the final predictions.

    \item \textbf{LSTM-MJT:} The LSTM-MJT model enhances the pure LSTM approach by incorporating additional inputs from the MJT. The time-series data is processed by the LSTM layer (as in the pure LSTM model), but additional inputs from the MJT—namely the predicted grasp position and time—are also included. These additional inputs are passed through separate dense layers, one with 16 units for processing the predicted position and another with 8 units for processing the predicted time. The outputs from the LSTM layer and the dense layers are then concatenated to produce the final predictions.

    The models were trained using the Adam optimizer with a learning rate of 0.001 for 50 epochs, with batch size set to 64. Time prediction was optimized using the Mean Absolute Error (MAE) loss function, while position prediction used the Mean Squared Error (MSE) loss function. To balance the importance of the two objectives, the losses were weighted in a 3:1 ratio. Results were given by 5-fold cross validation.

\end{itemize}
    
\end{enumerate}

\textbf{Inputs and Outputs:}

All models predict the final grasp position $\hat{\mathbf{P}}_{\text{grasp}}$ and the grasp time $\hat{T}_{\text{grasp}}$. The inputs vary slightly across the three models:

- MJT: Time-series data $\left\{\mathbf{P}\left(t_i\right), \Delta t_i\right\}_{i=1}^n$.

- LSTM: Same time-series data $\left\{\mathbf{P}\left(t_i\right), \Delta t_i\right\}_{i=1}^n$.

- LSTM-MJT: Same time-series data $\left\{\mathbf{P}\left(t_i\right), \Delta t_i\right\}_{i=1}^n$, along with the additional predicted grasp position $\hat{\mathbf{P}}_{\text{jerk}}$ and predicted grasp time $\hat{T}_{\text{jerk}}$ from the MJT.

\subsubsection{Experimental Results}
\begin{figure}[h]
    \centering
    \includegraphics[width=\linewidth]{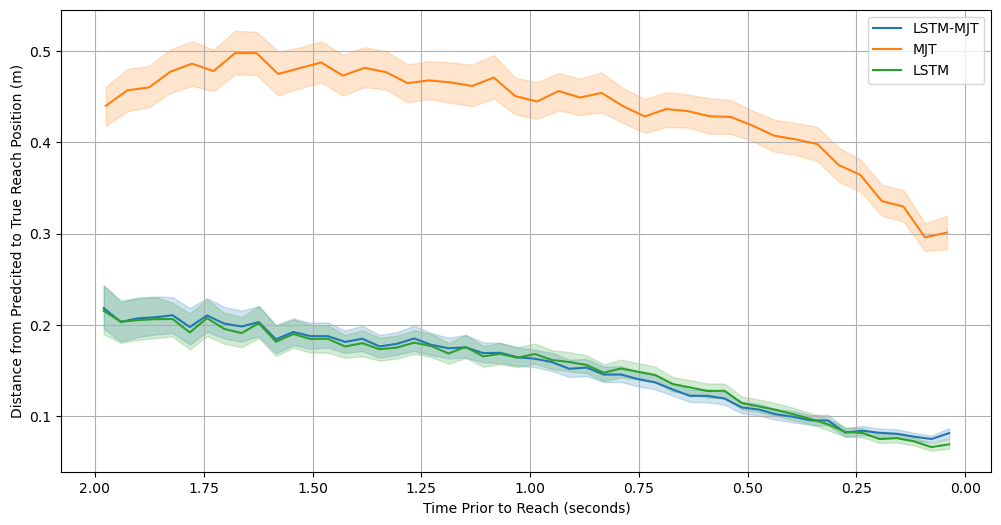}
    \caption{Distance from predicted to true reach position across different time points prior to reaching. Intervals show 95-CI.}
    \label{fig:distance}
\end{figure}

\begin{figure}[h]
    \centering
    \includegraphics[width=\linewidth]{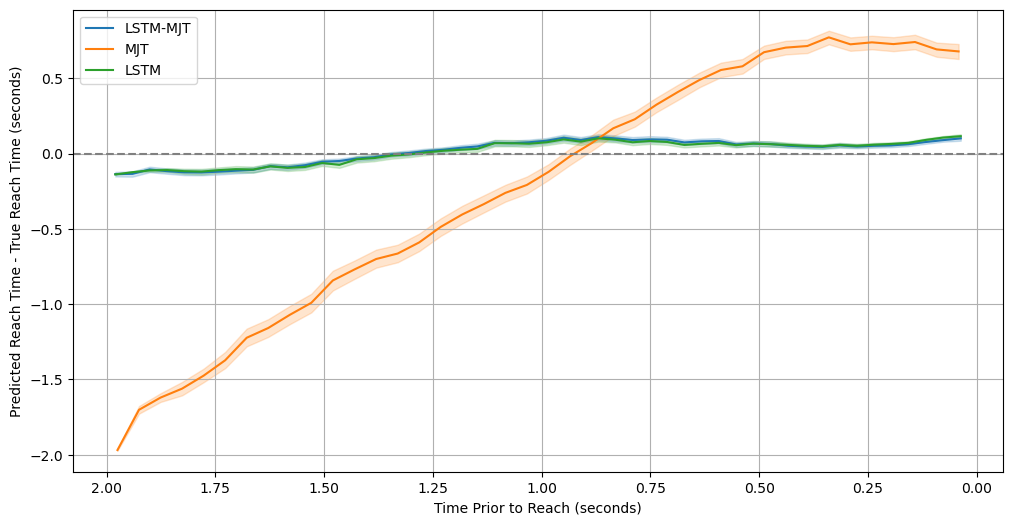}
    \caption{Predicted reach time error across different time points prior to reaching. Intervals show 95-CI.}
    \label{fig:time}
\end{figure}

In this section, we present and analyze the performance of three models: LSTM, MJT, and LSTM-Jerk. Figures \ref{fig:distance} and \ref{fig:time} illustrate the models' performance in predicting the final grasp position and the time to reach the grasp position, respectively, as a function of time prior to the actual reach.

LSTM and LSTM-MJT consistently outperform MJT in predicting the final grasp position (Figure \ref{fig:distance}) and time (Figure \ref{fig:time}), they achieve the lowest position error throughout the entire 2-second window, with a minimum error of approximately 0.18 meters occurring around 1 second before the grasp. This suggests that LSTM structure effectively captures the overall movement trajectory. However, in the final 0.25 seconds, both LSTM and LSTM-MJT exhibit a slight increase in error, indicating that predicting the final adjustments in hand approach remains challenging. This could be attributed to the complex, non-linear corrections users make in the final moments of the reach, which are difficult for the model to anticipate.

The MJT model, as expected, improves its predictions as the grasp event approaches, consistent with the assumption that closer data points yield better predictions. However, it consistently performs worse than the LSTM-based models, with a relatively high error (around 0.3 meters) even at the point of grasp. This suggests that while MJT can model idealized smooth trajectories, it struggles with the variability and complexity of real-world movements, where deviations from the minimum jerk assumption are common.

Interestingly, the LSTM-Jerk model, which combines the predictions of the MJT with the LSTM, does not show a significant improvement over the pure LSTM model. This indicates that incorporating biomechanical models like MJT into data-driven approaches does not always yield better results, particularly when the assumptions of the biomechanical model do not perfectly align with the real movement patterns.

In terms of time prediction (Figure \ref{fig:time}), all three models exhibit a similar pattern. Initially, predictions are early, underestimating the time required to reach the object. As more data becomes available, predictions improve, and all models cross the zero-error line, indicating closer alignment with the actual reach time. However, as more information is incorporated, predictions begin to lag behind the true grasp time. The early underestimation may be due to the models anticipating faster, more direct trajectories, whereas human movements often involve slight adjustments as the hand approaches the object. This shift to overestimation suggests that while the models can adapt, they struggle with maintaining consistent timing accuracy in the final moments before the grasp.

Overall, the slight struggles in the final moments of the grasp highlight the need for further refinement. Future work should continue refining LSTM-based approaches while carefully considering the trade-offs of integrating biomechanical models like MJT. Additionally, exploring alternative sequence modeling techniques such as Transformers could offer new insights and potentially enhance temporal prediction performance.

\subsubsection{Results Visualization} 
To illustrate the predicted trajectories of the LSTM, LSTM-Jerk, and MJT models over time, we randomly selected three users with specific configurations: User 3 (pull, cube, medium), User 1 (raise, 1, large), and User 7 (pull, cube, small). The figures below depict how each model's predictions evolve as the time to grasp decreases.

\begin{figure}[h] \centering \includegraphics[width=\linewidth]{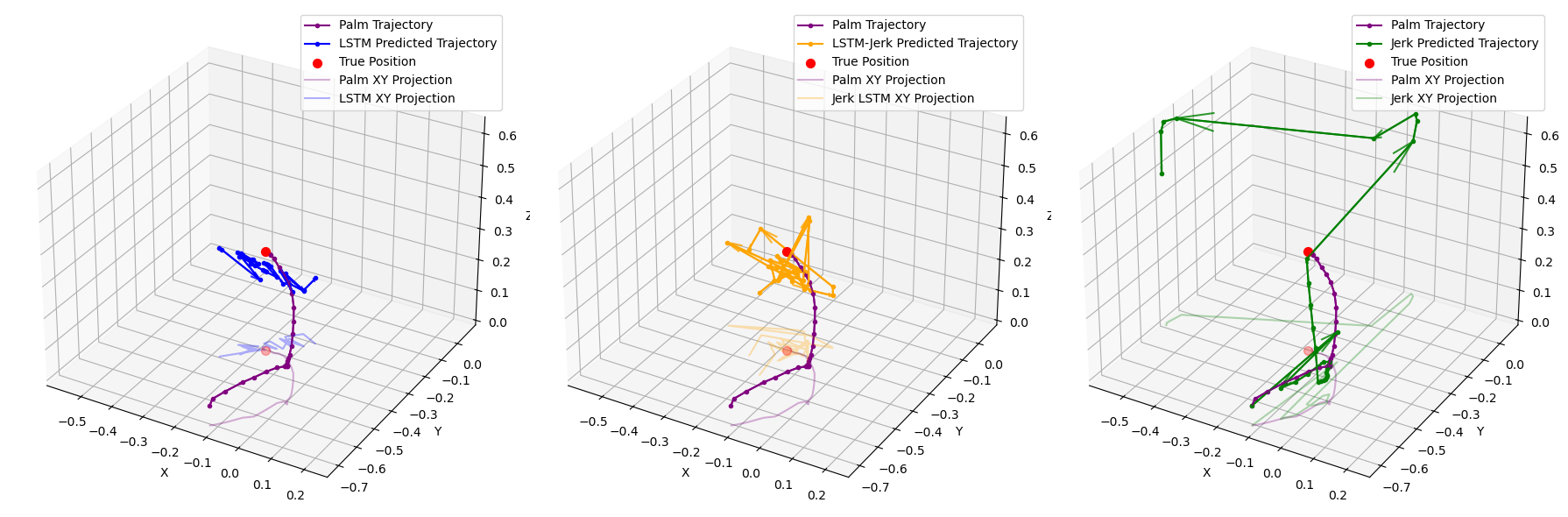} \caption{Trajectories of palm and predicted positions for User 3 (pull, cube, medium) across LSTM, LSTM-Jerk, and MJT models. The red dot indicates the true grasp position.} \label{fig
} \end{figure}

\begin{figure}[h] \centering \includegraphics[width=\linewidth]{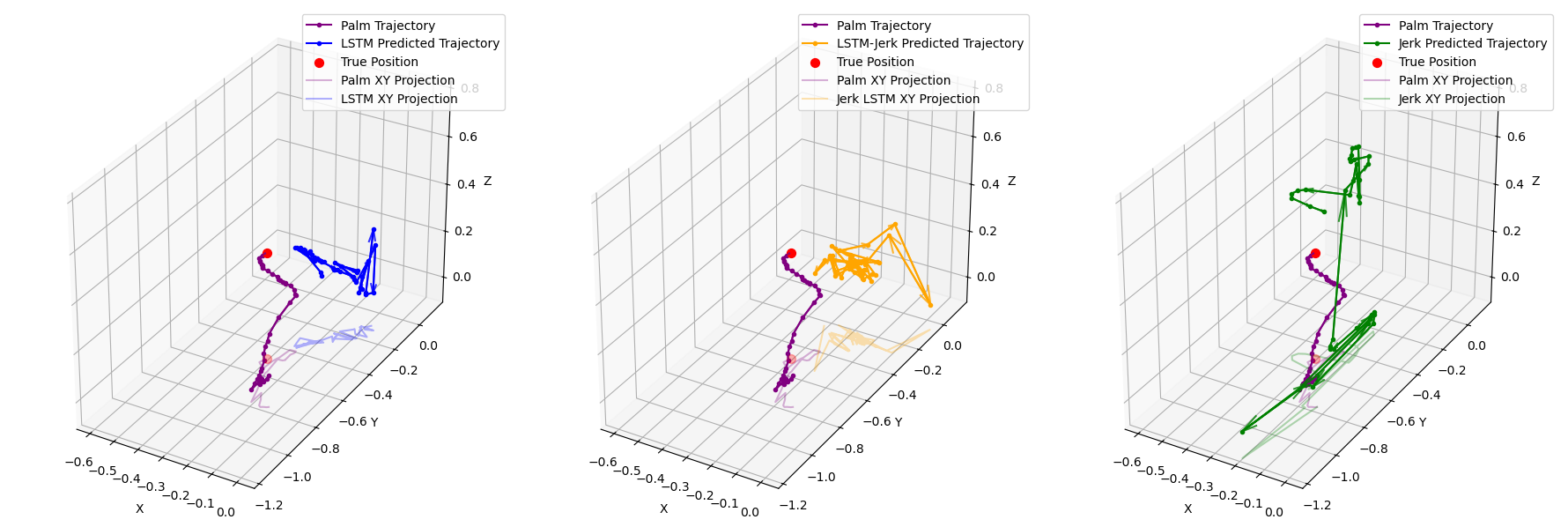} \caption{Trajectories of palm and predicted positions for User 1 (raise, 0, large) across LSTM, LSTM-Jerk, and MJT models. The red dot indicates the true grasp position.} \label{fig
} \end{figure}

\begin{figure}[h] \centering \includegraphics[width=\linewidth]{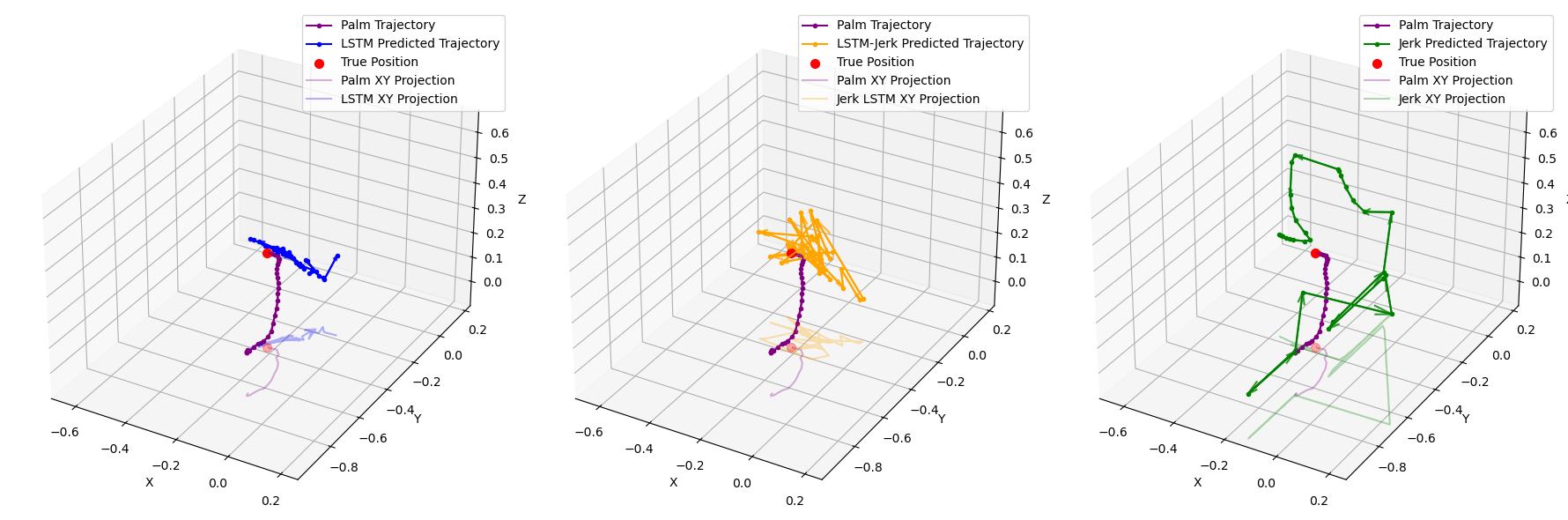} \caption{Trajectories of palm and predicted positions for User 7 (pull, cube, small) across LSTM, LSTM-Jerk, and MJT models. The red dot indicates the true grasp position.} \label{fig
} \end{figure}

The LSTM model, while achieving the lowest distance error, tends to keep its predictions within a consistent range around the true grasp point. This indicates that the model effectively minimizes error but may struggle to precisely converge on the exact position as the grasp event approaches. The LSTM-Jerk model, which combines the outputs of LSTM and MJT, shows more variability, with predictions oscillating around the true grasp point in a somewhat random manner. Although this variability suggests dynamic adjustment, it lacks the stability needed for consistent accuracy.

The MJT model, on the other hand, exhibits more significant fluctuations in its predictions, especially when small movements occur. This suggests that the MJT model is highly sensitive to minor variations in the hand’s trajectory, leading to greater prediction variability. Contrary to expectations, the MJT predictions do not consistently move towards the target location as the grasp event approaches, indicating that this model struggles to maintain a steady direction towards the grasp point.

These visualizations highlight the distinct strengths and limitations of each model. The LSTM excels at reducing distance error but may not always follow the most natural or precise trajectory. The LSTM-Jerk model, while more adaptive, introduces variability that can undermine stability. The MJT model, despite showing greater sensitivity to small movements, does not reliably guide the predictions toward the target, resulting in less predictable outcomes.

\subsection{Predicting the Posture of the Hand When Grasp Happens}
\subsubsection{Problem Setup}
\textbf{Input Data:}
For the hand posture prediction task, we continue to focus on right-hand data within the final 2-second window leading up to the grasp event. The key input data consists of vectors from the palm to the five fingertips, capturing the hand's configuration as it approaches the object. These five vectors represent the primary features used to describe the hand's posture, and they provide a simplified but effective representation of the overall hand shape. The aim is to predict the final configuration of these vectors at the moment of grasp.

The input data is structured as a time-series of 3D vectors, where each vector represents the position of a fingertip relative to the palm. Formally, for each fingertip $j$ at time step $i$, the vector can be expressed as:

$$
\vec{u}_j\left(t_i\right)=\left[\mathbf{F}_j\left(t_i\right)-\mathbf{P}\left(t_i\right)\right]
$$

where $\mathbf{F}_j(t_i)$ is the 3D coordinate of fingertip $j$ at time step $i$, and $\mathbf{P}(t_i)$ is the 3D coordinate of the palm at the same time step. In addition to these vectors, we also include the time differences between consecutive frames, denoted as $\Delta t_i=t_{i}-t_{i-1}$, to account for the uneven sampling rate of the tracking device. And to capture different stages of the hand's motion and explore when predictions are most effective, we again employ the sliding window technique as before. 

Formally, the input to the model for a sequence of $\$ \mathrm{n} \$$ time steps can be represented as:

$$
\mathbf{X}=\left\{\left[\vec{u}_1\left(t_1\right), \vec{u}_2\left(t_1\right), \ldots, \vec{u}_5\left(t_1\right), \Delta t_1\right], \ldots,\left[\vec{u}_1\left(t_n\right), \vec{u}_2\left(t_n\right), \ldots, \vec{u}_5\left(t_n\right), \Delta t_n\right]\right\}
$$

\textbf{Prediction Targets:}
The output of the model is a set of five 3D vectors corresponding to the predicted positions of the fingertips relative to the palm at the moment of grasp. These vectors describe the final posture of the hand and are denoted as:

$$
\mathbf{Y}=\left\{\vec{u}_1^{\text { grasp }}, \vec{v}_2^{\text { grasp }}, \vec{u}_3^{\text { grasp }}, \vec{u}_4^{\text { grasp }}, \vec{u}_5^{\text { grasp }}\right\}
$$

Thus, the model's task is to predict the final configuration of the hand at the grasp moment based on the input time-series data. This prediction is crucial for determining how the user intends to grasp the object, which, in turn, informs the appropriate haptic feedback to be provided during the interaction.

\subsubsection{Method}
We initially experimented with fixed-length data sequences for predicting hand posture at the moment of grasp. In this setup, we tested several machine learning models, including Linear Regression, Decision Tree, Random Forest, Support Vector Regression (SVR), and LSTM networks. The loss function for all models was the MSE. After evaluating these models, we found that while some achieved reasonable performance with fixed-length data, real-time prediction scenarios require models that can handle variable-length inputs. Given this need for flexibility, we selected the LSTM model for further experiments due to its ability to process sequential data effectively.

For our LSTM model, we used a single LSTM layer with 64 units, followed by a fully connected layer with 32 units. Both layers employed the ReLU activation function. To prevent overfitting, we included a dropout rate of 0.2 after the LSTM layer. The model was trained using the Adam optimizer with a learning rate of 0.001 for 100 epochs, and we used a batch size of 64 during training. The loss function remained MSE. Results were given by 5-fold cross validation.

After visualizing the model's predictions, we observed that the predictions appeared to be randomly fluctuating, which could negatively affect the user experience in a real-time application. To address this issue, we introduced an additional constraint to the LSTM architecture by adding a temporal smoothing loss function. This function ensures that the predictions at time step $t$ and $t+1$ do not deviate excessively, thereby promoting more consistent predictions over time.

The additional loss term can be defined as:
$$
\text { Temporal Smoothness Loss}=\lambda \cdot \sum_{i=1}^{n-1}\left\|\hat{\vec{u}}_i-\hat{\vec{u}}_{i+1}\right\|^2
$$

where $\hat{\vec{u}}_i$ represents the predicted fingertip vectors at time step $i$, and $\lambda$ is a regularization parameter that controls the strength of the smoothness constraint. This loss term was added to the primary MSE loss to create the final loss function for the LSTM model. By incorporating this temporal smoothing loss, we aimed to generate smoother, more stable predictions that would enhance the user experience in a real-time VR environment.

\subsubsection{Experimental Results}
\begin{figure}[h]
    \centering
    \includegraphics[width=\linewidth]{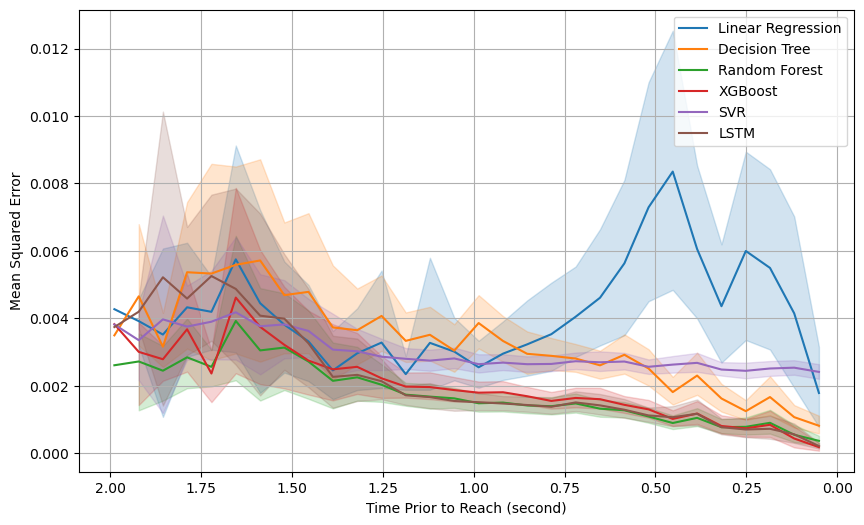}
    \caption{Mean Squared Error (MSE) of the predicted vectors across different time points prior to reaching.}
    \label{fig:posmse}
\end{figure}

In this section, we evaluate the performance of different machine learning models in predicting hand posture at the moment of grasp on, focusing on Mean Squared Error (MSE) and average Euclidean distance difference between the predicted and true vectors. All the results were given by 5-fold cross validation.

\begin{figure}[h]
    \centering
    \includegraphics[width=\linewidth]{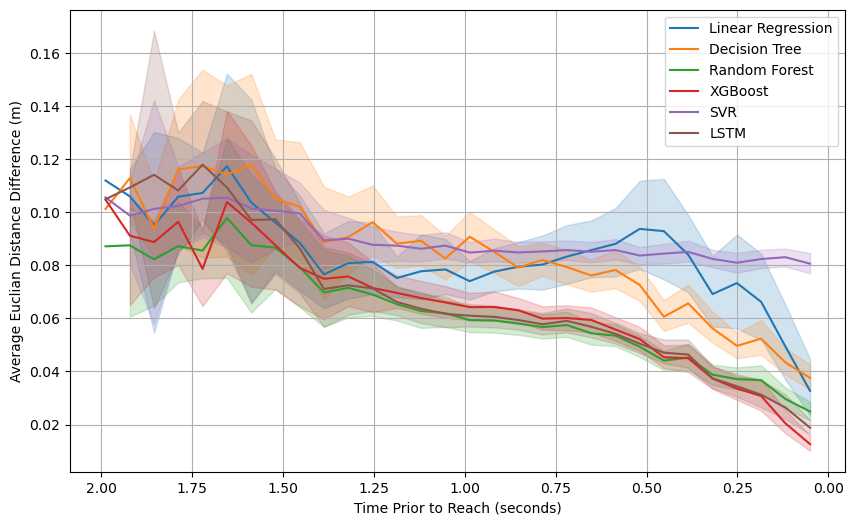}
    \caption{Average Euclidean distance difference between the predicted and true vectors across different time points prior to reaching.}
    \label{fig:posdis}
\end{figure}

\textbf{Fixed Length Data:}
As shown in Figure \ref{fig:posmse}, MSE generally decreases as the time to grasp approaches, indicating an improvement in prediction accuracy across all models with more available data. Among the models tested, Random Forest and XGBoost consistently demonstrate the best performance, with Random Forest achieving the lowest MSE across most of the 2-second window. The LSTM model also performs well, especially in the final moments before the grasp, though it does not outperform the tree-based models. As expected, Linear Regression exhibits the highest MSE, particularly in the last 0.5 seconds, reflecting its limitations in capturing the non-linear dynamics of hand movements.

Figure \ref{fig:posdis} illustrates the average Euclidean distance difference between the predicted and true vectors. Similar to MSE, the Euclidean distance difference decreases as the time of grasp approaches, with Random Forest consistently maintaining the lowest distance error. Interestingly, while Linear Regression shows the highest MSE, its Euclidean distance error is sometimes lower than that of Decision Tree and SVR, suggesting that MSE alone may not fully capture the visual accuracy of predictions.

 While the fixed-length input approach works well for controlled experiments, it may not be well-suited for dynamic, real-time environments where input data lengths vary. Tree-based models like Random Forest and XGBoost perform well in this controlled setting, achieving low MSE and Euclidean distance errors. However, their reliance on fixed-length inputs could hinder their effectiveness in more dynamic scenarios. On the other hand, LSTM models, with their ability to handle variable sequence lengths, maintain competitive performance and offer more flexibility for real-time applications. This makes LSTM a promising candidate for prediction tasks in VR. As a result, our subsequent experiments focused on refining LSTM for use with variable-length time-series data to ensure it can handle the demands of real-world, dynamic VR interactions.

\textbf{Variable Length Data:}

\begin{figure}[h] 
\centering \includegraphics[width=\linewidth]{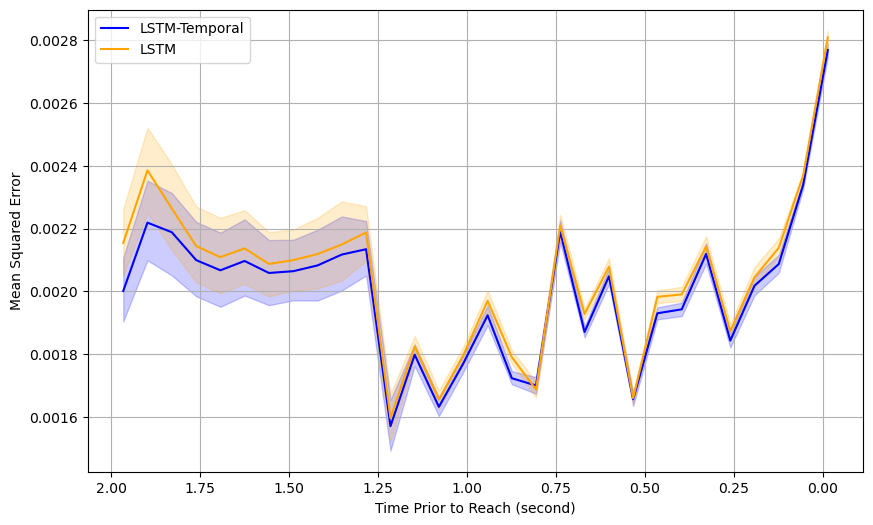} \caption{MSE of the predicted hand posture vectors across different time points prior to reaching.} 
\label{fig:mse-v} 
\end{figure}

\begin{figure}[h] \centering \includegraphics[width=\linewidth]{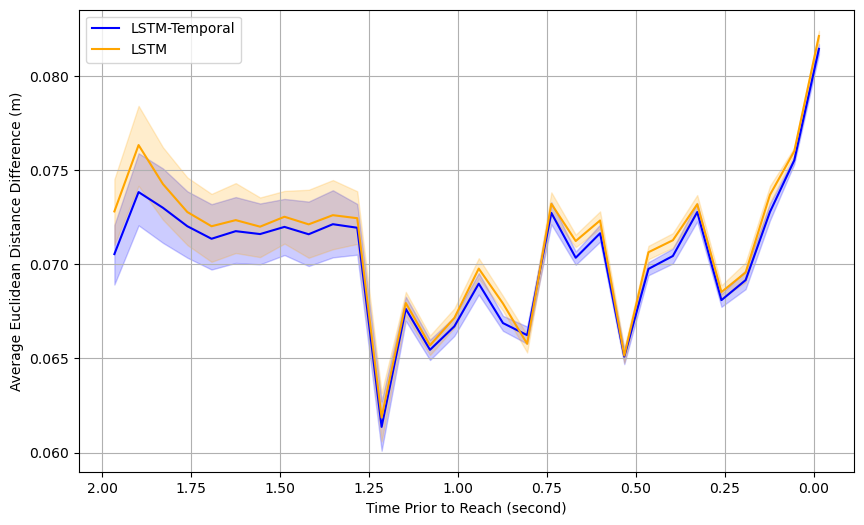} \caption{Average Euclidean distance difference between the predicted and true hand posture vectors across different time points prior to reaching.} \label{fig:dist-v} \end{figure}

After implementing the temporal smoothness constraint, we evaluated the performance of the LSTM model against the original LSTM configuration using variable-length time-series data. The two models were compared in terms of MSE and average Euclidean distance difference across different time points leading up to the grasp.

As illustrated in Figures \ref{fig:mse-v} and \ref{fig:dist-v}, the LSTM model with the temporal smoothing constraint (LSTM-Temporal) generally shows a slight improvement over the standard LSTM in both MSE and Euclidean distance difference. Both models exhibit a similar trend where the error decreases during the first second as the time to grasp decreases, followed by some fluctuations, and finally a sharp increase in the last 0.25 seconds before the grasp. This sharp increase suggests that predicting the exact posture in the final moments remains challenging, likely due to the rapid adjustments in hand movements. This highlights the challenge of achieving stable and accurate predictions in real-time VR applications where rapid movements are involved.

\begin{figure}[h]
    \centering
    \begin{subfigure}[b]{\textwidth}
        \centering
        \includegraphics[width=\textwidth]{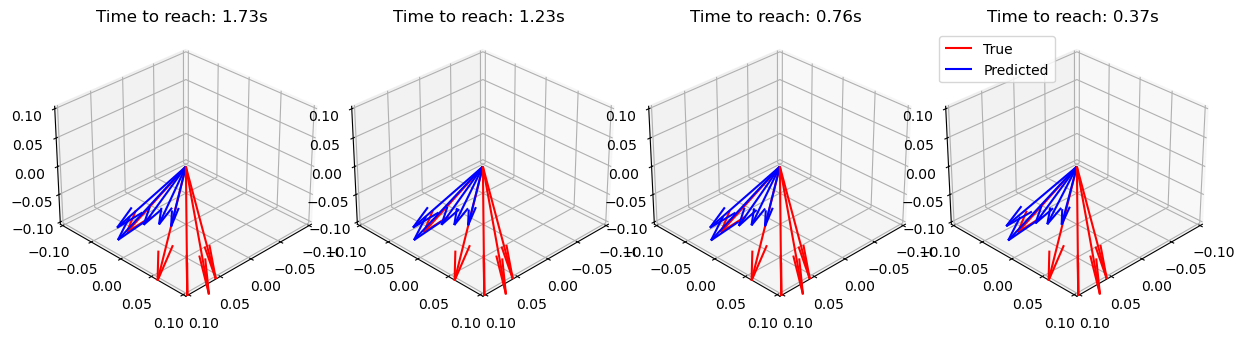}
        \caption{LSTM-Temporal for User 3, Pull, Cube, Medium}
    \end{subfigure}
    
    \begin{subfigure}[b]{\textwidth}
        \centering
        \includegraphics[width=\textwidth]{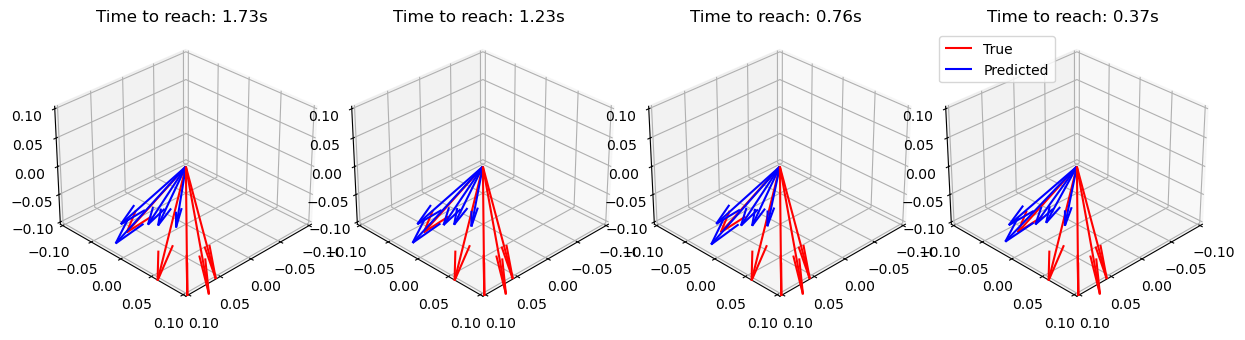}
        \caption{LSTM for User 3, Pull, Cube, Medium}
    \end{subfigure}
    
    \begin{subfigure}[b]{\textwidth}
        \centering
        \includegraphics[width=\textwidth]{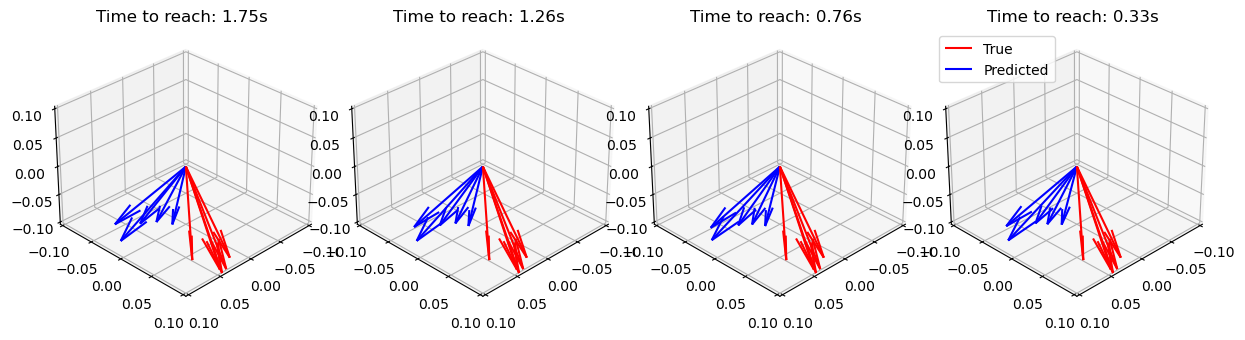}
        \caption{LSTM-Temporal for User 1, Touch, Small Object}
    \end{subfigure}
    
    \begin{subfigure}[b]{\textwidth}
        \centering
        \includegraphics[width=\textwidth]{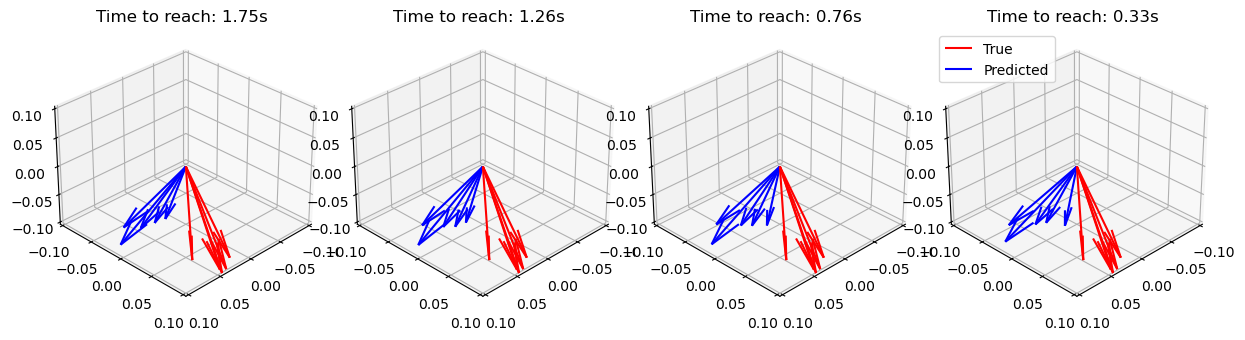}
        \caption{LSTM for User 1, Touch, Small Object}
    \end{subfigure}
    
    \caption{Comparison of predicted final posture with and without temporal smoothing for different users and actions. Red represents the true posture, and blue represents the predicted posture.}
    \label{fig:temp_vs_non_temp}
\end{figure}

\subsubsection{Results Visualization}

While the quantitative analysis provided by the MSE and average Euclidean distance difference metrics offers valuable insights, it may not fully capture the qualitative impact of the temporal smoothing constraint on the prediction trajectories. Specifically, these metrics alone do not reveal whether the LSTM-Temporal model truly generates smoother, more consistent predictions over time.

To visually compare the effects of temporal smoothing on the LSTM model's predictions, we selected two specific cases: User 3 performing a pull action on a medium cube and User 1 performing a touch action on a small object. These visualizations are shown in Figure \ref{fig:temp_vs_non_temp}, where we compare the grasp posture predicted by the LSTM with and without the temporal constraint.

From the visualizations in Figure \ref{fig
}, it is difficult to see significant differences between the predictions of the LSTM model with temporal smoothing and the standard LSTM model. This may be due to the relatively small differences in MSE between the two models (less than 0.001), making it hard to observe improvements in trajectory smoothness or consistency. 

The lack of precision improvement over time might be due to the model's training process, where sequences of different lengths (e.g., [0,1] seconds and [0,2] seconds) were trained to predict the same final grasp position. This could lead the model to rely primarily on the early part of the sequence, underutilizing the additional data from longer sequences. As a result, the model's predictions may remain stable across different sequence lengths but do not show significant improvements in accuracy or smoothness as the sequence length increases. This behavior suggests that the model may "ignore" later parts of the sequence, not fully utilizing the available temporal information. To address this, varying the training targets or adjusting the training strategy—such as incorporating intermediate targets or applying a weighted loss function—could help the model make better use of longer sequences, leading to more accurate and consistent predictions. Additionally, increasing the $\lambda$ value in the temporal smoothing constraint might enhance prediction stability, especially in real-time VR applications.

\subsection{Discussion}
The analysis of our experimental results show several advantages of using regression over classification for predicting user intentions in VR environments. These insights suggest that regression-based approaches are more aligned with the dynamic and continuous nature of human behavior, offering more practical solutions for real-time applications.

\begin{enumerate}
    \item Enhanced Precision through Continuous Prediction: One of the primary advantages of regression-based models is their ability to predict continuous outcomes, such as the exact position, posture, and time of grasp. Unlike classification, which forces discrete categorization, regression allows for more nuanced predictions that can closely match the variability and subtlety of human movements.
    \item Challenges in Integration with Biomechanical Models: While integrating biomechanical insights like the MJT into regression models provided some benefits, it also introduced challenges. The combined LSTM-Jerk model, for example, showed more variability in its predictions, which could hinder its effectiveness in real-time scenarios. This calls the need for careful consideration when combining data-driven approaches with biomechanical models, ensuring that the integration enhances rather than detracts from the model’s predictive capabilities.
    \item Scalability and Real-Time Application: Regression models, especially LSTM-based ones, have demonstrated their scalability to handle variable-length input sequences, making them more suitable for dynamic, real-time VR environments. This scalability is a significant advantage over classification-based methods, which often struggle with the variability and complexity of real-world data.
    \item Training Strategy for Short and Long Sequences: An assumption from our study is that the model's training strategy might have influenced prediction precision over time. Specifically, when training sequences of different lengths (e.g., [0,1] seconds and [0,2] seconds) were used to predict the same final grasp position, the model appeared to focus primarily on the early part of the sequence. This could lead to underutilization of additional data provided by longer sequences, resulting in predictions that are relatively stable but do not significantly improve as the sequence length increases. It may be beneficial to explore varying the training targets or incorporating intermediate prediction goals in future work. Adjusting the training process in this way could potentially enable the model to make more effective use of longer sequences, thereby improving both accuracy and consistency.
\end{enumerate}
 
In conclusion, regression-based approaches seem to offer a more viable and effective framework for predicting user intentions in VR, particularly in tasks involving continuous and dynamic human movements. The flexibility, precision, and adaptability of regression models make them promising candidates for addressing the challenges of real-time VR interactions. However, further advancing these models to achieve even higher performance is still a challenge.

\section{Discussion and Future Work}
The findings from our study highlight the limitations of classification-based approaches and the advantages of regression-based methods for predicting user intentions in VR environments. 

\begin{enumerate}
    \item Regression-Based Approaches
    \begin{itemize}
        \item Advanced Models: Our study demonstrates that deep learning models, particularly LSTM, offer significant advantages in handling the temporal dynamics of user movements. However, integrating biomechanical principles like the Minimum Jerk theory with neural networks remains challenging. Future work should focus on refining hybrid models that can combine the strengths of these approaches, potentially leading to more accurate and stable predictions. These advanced models could better capture the subtleties of human motion, improving the reliability of predictions in VR applications.
        \item Multi-Modal Integration: Another promising direction for future research is the integration of additional data sources, such as eye tracking and environmental context. Multi-modal data could provide a more comprehensive understanding of user intentions, leading to more accurate and robust predictions. For instance, incorporating eye-tracking data could help anticipate where the user intends to interact, enhancing the precision of hand posture predictions \cite{BINSTED2001563}.
    \end{itemize}
    \item Improved Data Collection
    \begin{itemize}
        \item Structured Data Collection: The variability in user behavior observed in our study underscores the need for more systematic and controlled data collection. Future experiments should be designed to capture a wide range of user interactions in a controlled environment, ensuring that the data collected is both comprehensive and representative of real-world VR scenarios.

        \item Diverse and Larger Datasets: To improve the generalizability of our models, it is crucial to increase the diversity of the datasets. This can be achieved by involving a broader participant base and including a wider range of tasks. A more diverse dataset would help models better account for individual differences in behavior, leading to more reliable predictions across different users and contexts.
    \end{itemize}
    \item Implications for HCI and VR
    \begin{itemize}
        \item Adaptive Haptic Feedback: Effective haptic feedback relies on accurately predicting user intentions, but the inherent variability in user behavior makes this a challenging task. Future systems could benefit from adaptive prediction models that dynamically adjust feedback based on real-time user interactions. This approach would enhance the realism and immersion of VR experiences, allowing users to interact with virtual environments in a more natural and intuitive way.
    \end{itemize}
    \item Implications for Machine Learning
    \begin{itemize}
        \item Generalization and Adaptability: The development of regression-based models that generalize well across different users and tasks is crucial for their success in VR applications. Future work should explore techniques such as transfer learning to improve the generalization capabilities of these models. Additionally, semi-supervised learning approaches could be investigated to handle user variability without requiring extensive labeled data, which is often challenging to obtain.
        \item Real-Time Performance: As VR applications demand real-time processing, it is essential to prioritize the development of computationally efficient models. Future research should focus on optimizing the performance of regression models to ensure they can operate within the constraints of real-time VR environments. This could involve exploring lightweight architectures or leveraging hardware accelerations to achieve the necessary computational efficiency.
    \end{itemize}

\end{enumerate}

\section{Conclusion}
In this work, we explored the challenges and potential solutions for predicting user intentions in virtual environments, focusing on the tasks of predicting the position, timing, and posture of grasps. Our study revealed significant challenges in using classification models, particularly their difficulty in generalizing across different users. This limitation underscores the complexity of user behavior in VR, which often leads to inconsistent performance when using classification approaches.

Conversely, our regression-based approach, specifically LSTM-based models, showed more promise. Within the two seconds leading up to a grasp, the LSTM model maintained a prediction error of less than 0.25 seconds for timing and around 5-20 cm for distance. However, predicting the precise hand posture remains a challenge, with relatively large errors observed in this aspect.

These findings highlight the potential of regression models in offering more adaptable and accurate predictions in dynamic VR environments, while also pointing out the areas where further improvement is needed, particularly in predicting complex hand postures. Our work lays the foundation for future research aimed at refining these models and better understanding the intricacies of user haptic interactions in VR.

\section*{Acknowledgements}
\vspace{-1em} 
We sincerely thank Elodie Bouzbib for conducting the data collection and Clement Alberge for his support in configuring the VR software.

\bibliographystyle{ieeetr}  
\bibliography{sources}


\end{document}

%% file: config/settings.tex
\definecolor{mpblue}{HTML}{005E9E}

\titleformat{\chapter}
{\color{BlueViolet}\normalfont\huge\bfseries}
{\color{BlueViolet}\thechapter.}{1em}{}
\titlespacing*{\chapter}
{0pt}{3.3ex}{3.3ex}

\titleformat{\section}
{\color{BlueViolet}\normalfont\Large\bfseries}
{\color{BlueViolet}\thesection.}{1em}{}
\titlespacing*{\section}
{0pt}{3.3ex}{3.3ex}

\titleformat{\subsection}
{\color{BlueViolet}\normalfont\large\bfseries}
{\color{BlueViolet}\thesubsection.}{1em}{}
\titlespacing*{\subsection}
{0pt}{3.3ex}{3.3ex}

\titleformat{\subsubsection}
{\color{BlueViolet}\normalfont\large\bfseries}
{\color{BlueViolet}\thesubsubsection.}{1em}{}
\titlespacing*{\subsubsection}
{0pt}{3.3ex}{3.3ex}

\declaretheoremstyle[
  headfont=\color{BlueViolet}\normalfont\bfseries,
  bodyfont=\color{black}\normalfont\itshape,
]{colored}

\theoremstyle{colored}

\makeatletter
\newcommand\subtitle[1]{\renewcommand\@subtitle{#1}}
\newcommand\@subtitle{}
\makeatother

\makeatletter
\newcommand\keywords[1]{\renewcommand\@keywords{#1}}
\newcommand\@keywords{}
\makeatother

\makeatletter
\newcommand\logo[1]{\renewcommand\@logo{#1}}
\newcommand\@logo{}
\makeatother

\makeatletter
\newcommand\secondlogo[1]{\renewcommand\@secondlogo{#1}}
\newcommand\@secondlogo{None}
\makeatother

\newcommand\none{None}
\newcommand\includegraphicsifexists[2]{\IfFileExists{#2}{\includegraphics[#1]{#2}}}

\makeatletter
\newcommand\firstsupervisor[1]{\renewcommand\@firstsupervisor{#1}}
\newcommand\@firstsupervisor{None}
\makeatother

\makeatletter
\newcommand\secondsupervisor[1]{\renewcommand\@secondsupervisor{#1}}
\newcommand\@secondsupervisor{None}
\makeatother

\makeatletter
\newcommand\thirdsupervisor[1]{\renewcommand\@thirdsupervisor{#1}}
\newcommand\@thirdsupervisor{None}
\makeatother

\makeatletter
\newcommand\fourthsupervisor[1]{\renewcommand\@fourthsupervisor{#1}}
\newcommand\@fourthsupervisor{None}
\makeatother

\makeatletter
\newcommand\fifthsupervisor[1]{\renewcommand\@fifthsupervisor{#1}}
\newcommand\@fifthsupervisor{None}
\makeatother

\makeatletter
\newcommand\firstsupervisorrole[1]{\renewcommand\@firstsupervisorrole{#1}}
\newcommand\@firstsupervisorrole{}
\makeatother

\makeatletter
\newcommand\secondsupervisorrole[1]{\renewcommand\@secondsupervisorrole{#1}}
\newcommand\@secondsupervisorrole{}
\makeatother

\makeatletter
\newcommand\thirdsupervisorrole[1]{\renewcommand\@thirdsupervisorrole{#1}}
\newcommand\@thirdsupervisorrole{}
\makeatother

\makeatletter
\newcommand\fourthsupervisorrole[1]{\renewcommand\@fourthsupervisorrole{#1}}
\newcommand\@fourthsupervisorrole{}
\makeatother

\makeatletter
\newcommand\fifthsupervisorrole[1]{\renewcommand\@fifthsupervisorrole{#1}}
\newcommand\@fifthsupervisorrole{}
\makeatother

\makeatletter
\renewcommand{\maketitle}{
    {\vspace*{2.8cm}}
    {\centering}
    {\LARGE{\textbf{\@title}}}
    
    {\vspace*{0.25cm}}
    {\large {\textbf{\@subtitle}}}
    
    {\vspace{3cm}}
    {\large{\textbf{Author: \@author}}}
    
    {\vspace{2cm}}
    {\large{\textbf{\@date}}}
}
\makeatother

%% file: config/titlepage.tex
\begin{titlepage}
    \begin{center}
    \makeatletter 

        \begin{figure}[!htb]
        \centering
        \begin{minipage}{0.5\textwidth}
            \centering       
            \includegraphics[width=\linewidth]{\@logo}
        \end{minipage}
        \hspace{1cm}
        \end{figure}

    \makeatother
    \vspace*{0.25cm}

    \maketitle

    \vspace{2cm}
    \makeatletter
    \ifx\@secondlogo\none
    \else
        \textbf{Hosting Institute:}
        \vspace{0.5cm}
        \includegraphics[width=0.2\textwidth]{\@secondlogo}\\
        \textit{Institut des systèmes intelligents et de robotique}
        \vspace{1cm}
    \fi
    \makeatother

    \vspace*{\fill}
    \normalsize
    Conducted under the supervision of: \\
    \vspace{0.5cm}
    \centering
    \makeatletter
    \begin{tabular}{l c r}
    \ifx\@firstsupervisor\none  \else \hspace{0.5cm}  \@firstsupervisor       & \hspace{0.25cm}-\hspace{0.25cm} &  \@firstsupervisorrole \\ \fi 
    \ifx\@secondsupervisor\none \else \hspace{0.5cm}  \@secondsupervisor      & \hspace{0.25cm}-\hspace{0.25cm} &  \@secondsupervisorrole \\ \fi
    \ifx\@thirdsupervisor\none \else \hspace{0.5cm}  \@thirdsupervisor       & \hspace{0.25cm}-\hspace{0.25cm} &  \@thirdsupervisorrole \\ \fi 
    \ifx\@fourthsupervisor\none \else \hspace{0.5cm}  \@fourthsupervisor      & \hspace{0.25cm}-\hspace{0.25cm} &  \@fourthsupervisorrole \\ \fi 
    \ifx\@fifthsupervisor\none \else \hspace{0.5cm}  \@fifthsupervisor       & \hspace{0.25cm}-\hspace{0.25cm} &  \@fifthsupervisorrole \\ \fi 
    \end{tabular}
    \makeatother

    \end{center}
\end{titlepage}

\thispagestyle{empty}
\cleardoublepage